\documentclass[aps,amsmath,amssymb,twocolumn,footinbib,prl]{revtex4-1}

\usepackage{graphicx}
\usepackage{color}
\usepackage{soul}
\usepackage{lmodern} 
\usepackage{lmodern}
\usepackage{amsmath}
\usepackage{amsthm}
\usepackage{amsfonts}
\usepackage{natbib}
\usepackage{placeins}

\newcommand{\bra}[1]{\ensuremath{\left\langle #1 \right|}}
\newcommand{\ket}[1]{\ensuremath{\left| #1 \right\rangle}}

\newcommand{\ketbra}[2]{\ensuremath{\left|#1\right\rangle \left\langle#2\right|}}

\newcommand{\Sij}[3]{ \left \langle #1 \right | #2 \left | #3 \right \rangle}

\newcommand{\bd}{\ensuremath{b^\dagger}}

\newcommand{\XM}{\ensuremath{X}}

\newcommand{\nbar}{\ensuremath{\bar{n}}}

\newcommand{\omegam}{\ensuremath{\omega_{\textrm{m}}}}
\newcommand{\omegaq}{\ensuremath{\omega_{\textrm{q}}}}

\newcommand{\makesub}[1]{\textrm{\scriptsize{#1}}}

\newcommand{\Exp}[1]{\ensuremath{\exp\left ( #1\right)}}

\newcommand{\DD}[2]{\ensuremath{\frac{d #1}{d #2}}}

\newcommand{\sigmaz}{\ensuremath{\sigma_{\textrm{z}}}}

\newcommand{\sigmap}{\ensuremath{\sigma_{\textrm{+}}}}
\newcommand{\sigmam}{\ensuremath{\sigma_{\textrm{--}}}}

\begin{document}

\title{Arbitrary state preparation of a mechanical resonator via controlled pulse shaping and projective measurement in a qubit-resonator interaction}
\author{Kiran E. Khosla$^{1,2}$}
\email{k.khosla@imperial.ac.uk}
\affiliation{$^1$QOLS, Blackett Laboratory, Imperial College London, London SW7 2AZ, United Kingdom.\\
$^2$Center for Engineered Quantum Systems, University of Queensland,
St Lucia 4072, Australia.}

\pacs{1}

\begin{abstract}
We introduce a protocol capable of generating a general measurement operator for a mechanical resonator. The technique requires a qubit-resonator interaction and uses a coherent pulse to drive qubit transitions. This is followed by projective measurement of the qubit's energy, constraining the resonator in a state that depends on the pulse shape. The freedom to choose a pulse shape for the coherent drive enables an arbitrary position-basis measurement operator. Using this measurement operator, we outline a two pulse protocol that probabilistically generates a pure mechanical state with a desired wavefunction, with near unit fidelity for realizable parameters. 
\end{abstract}
\maketitle

\textit{Introduction} --- Quantum mechanics is an incredibly successful theory at predicting the dynamics of microscopic systems. However, in our macroscopic world we do not perceive and quantum behavior begging the question -- how large an object can be put in an exotic quantum state? Advances in technology have enabled observation of quantum effects in mechanical oscillators with macroscopic numbers of constituent particles~\cite{aspelmeyer_cavity_2014}. State of the art opto and electromechanical devices implement a rich variety of designs, and allow coupling between qubits, fields and mechanical motion~\cite{pirkkalainen_hybrid_2013,lecocq_resolving_2015,pirkkalainen_cavity_2015,heikkila_enhancing_2014,sillanpaa_accessing_2009}, paving the way for strong coupling, and investigating quantum effects in mechanical systems~\cite{teufel_circuit_2011,armour_probing_2008,armour_entanglement_2002,blencowe_probing_2008,blencowe_nanoelectromechanical_2005,safavi-naeini_observation_2012}. In particular, experimental results have demonstrated ground state cooling~\cite{oconnell_quantum_2010}, squeezing~\cite{wollman_quantum_2015}, entanglement~\cite{palomaki_entangling_2013,riedinger_non-classical_2016}, and coherent control~\cite{kerckhoff_tunable_2013,lee_cooling_2010}, all with the motional state of a mechanically compliment element. 

Mechanical oscillators are ideal systems to probe quantum mechanics at ever larger mass scales, potentially testing macro realism~\cite{asadian_probing_2014}, the superposition principle~\cite{khosla_displacemon_2018,marshall_towards_2003}, and modifications to quantum mechanics~\cite{pikovski_probing_2012,li_detecting_2017,nimmrichter_optomechanical_2014,bassi_gravitational_2017}. Similar mesoscopic resonators are also strong candidates for quantum information applications such as error corrected quantum memories~\cite{wallquist_hybrid_2009,didier_quantum_2014,gilchrist_schrodinger_2004,leghtas_hardware-efficient_2013}, microwave-to-optical photon conversion~\cite{regal_cavity_2011}, and quantum meteorology~\cite{harris_feedback-enhanced_2012,khosla_quantum_2017,forstner_ultrasensitive_2014,toscano_sub-planck_2006}. Many of these applications require exotic quantum states for optimal operation, and while there are some existing proposals for generating quantum states of mechanical oscillators~\cite{palomaki_coherent_2013,bennett_quantum_2016}, universal wavefunction shaping, and state generation has remained absent. 

In this work we introduce a protocol capable of realizing an arbitrary quantum state via wavefunction shaping requiring only classical drive and projective qubit measurement. We begin by introducing a weak quantum measurement procedure~\cite{caves_quantum-mechanical_1987} to constrain the position probability distribution of a mechanical oscillator, and show arbitrary constraints are realizable. We then derive the required conditions for the constrained probability to result in an arbitrary wavefunction, and outline a two-step measurement protocol capable of achieving this. Finally we discuss realistic parameters demonstrating dissipative evolution can be safely neglected with experimentally feasible devices.

\textit{Model} --- Our model consists of a qubit coupled to a mechanical oscillator with resonance frequency $\omegam$. The coupling arises from a displacement-dependent qubit frequency, $\omegaq(\XM)$. The system Hamiltonian is $H/\hbar = \omegam \bd b + \frac12\omegaq(\XM) \sigmaz$, where $b$ is the mechanical annihilation operator, $\sqrt{2}\XM = b + \bd$ is the dimensionless position operator and $\sigmaz$ is the Pauli-z operator. For small displacements, the interaction is simplified by considering the frequency dependence to be well approximated by a perturbation $\omegaq(\XM)\sigmaz \approx \omegaq^{0}\sigmaz + 2\lambda_0 \XM\sigmaz$, where $2\lambda_0 = \DD{\omegaq}{\XM}$. This linearized interaction is well studied and has been observed both in the optical and microwave domains~\cite{heikkila_enhancing_2014,asadian_probing_2014,sillanpaa_accessing_2009,schneider_coupling_2012,cleuziou_carbon_2006,arcizet_single_2011,khosla_displacemon_2018}. In addition to the qubit-oscillator coupling we consider coherently driving the qubit with a time dependent pulse $\alpha(t)$. 

Moving into the interaction picture with $H_0/\hbar = \omegam\bd b + \frac12\omegaq^{0} \sigmaz$, the interaction Hamiltonian is
\begin{eqnarray}
H_I/\hbar = \lambda_0\XM\sigmaz + \alpha^*(t)\sigmap + \alpha(t) \sigmam,
\label{eq:Hamiltonianzz}
\end{eqnarray}
where $\sigmam$ ($\sigmap$) is the standard two level lowering (raising) operator, and the carrier frequency of $\alpha(t)$ is taken to be on resonance with the mean qubit frequency $\omegaq^{0}$ (zero detuning). Note this is a non-linear Hamiltonian as the Pauli z operator is itself quadratic in the two level raising and lowering operators.  The architecture considered in Ref.~\cite{khosla_displacemon_2018}, shows strong coupling in this Hamiltonian is achievable with current technology, and also describes how to dynamically toggle $\lambda_0$ on and off. For the moment we assume unitary evolution, and validate this approximation in subsequent sections. 


The interaction Hamiltonian, Eq.~\eqref{eq:Hamiltonianzz}, generates the following equations of motion,
\begin{subequations}
\begin{eqnarray}
\dot{c_e}(x,t) &=& -i{\lambda}_0x c_e(x,t) - i \alpha^*(t)c_g(x,t) \label{eq:ODEa} \\
\dot{c_g}(x,t) &=& ~~i{\lambda}_0x c_g(x,t) - i \alpha(t)c_e (x,t),
\label{eq:ODEb}
\end{eqnarray}\label{eq:ODE}
\end{subequations}
\!\!where $c_{e(g)}(x,t)$ is the probability amplitude of finding the qubit in the excited (ground) state, \emph{and} the oscillator at position $x$ at time $t$. If the qubit is measured and found to be in the excited state, the conditional state of the mechanical oscillator is $\ket{\psi(t)} \sim \int c_e(x,t)\ket{x}dx$, where $\sim$ denotes the fact that the right hand side is unnormalised. The unnormalized wavefunction is unsurprising as it is the remaining state after a projective measurement. Such a measurement constrains the state of the oscillator, only allowing components compatible with the qubit being in the excited state. The post-measurement wavefunction of the mechanical resonator is therefore proportional to $c_e(x,t)$, and herein lies the key to the protocol: different drive amplitudes $\alpha(t)$ can be used to engineer different wavefunctions of the mechanical resonator. This scheme builds on previous work using a measurement as a tool for state preparation~\cite{brawley_nonlinear_2016,brune_manipulation_1992,opatrny_improvement_2000}

There are existing proposals to generate exotic quantum states of motion using similar devices.
However these proposals require one to first generate a quantum state of the electromagnetic field~\cite{reed_faithful_2017}, require a different interaction Hamiltonian~\cite{khosla_quantum_2013,palomaki_coherent_2013,hofheinz_synthesizing_2009}, or both~\cite{berdova_micromanipulation_2013,vogel_quantum_1993,makhlin_quantum-state_2001}. In the current scheme $\alpha(t)$ is entirely classical negating the requirement quantum state engineering of the microwave field. Instead this proposal relies on the interference generated by the non-linear Hamiltonian, and projective measurement to generate an arbitrary motional state. 

\begin{figure}
	\centering
	\includegraphics[width=\columnwidth]{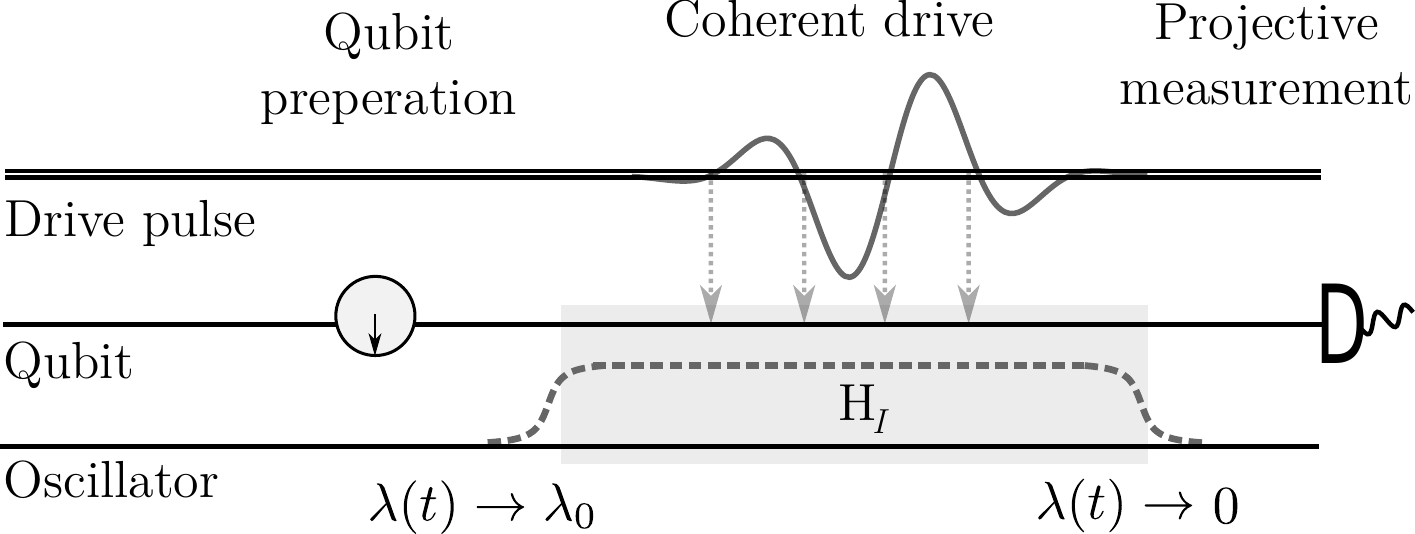}
	\centering
	\caption{Circuit diagram for the measurement protocol. The qubit is initially prepared in the $\sigmaz$ ground state with the qubit-oscillator interaction switched off. The interaction is then switched on and remains constant over the duration of the classical drive tone. At the completion of the drive tone, the interaction is once again switched off and the qubit measured in the $\sigmaz$ basis, conditioning the state of the oscillator.}
	\label{Fig_Schematic}
\end{figure}

We consider a measurement protocol as outlined in Fig.~\ref{Fig_Schematic}. The qubit is initialized in the ground state at $t = 0$ with the qubit-oscillator interaction switched off. The interaction is then switched on and remains constant over the duration of the drive pulse $\alpha(t)$. On the competition of the drive pulse, the qubit-oscillator interaction is again switched off, and a projective $\sigmaz$ measurement is made. The protocol lasts for a fixed duration $\tau$, understood to be the duration of the pulse $\alpha(t)$. This is the typical formulation of weak measurement, with the qubit as the measurement device~\cite{wiseman_quantum_2009,caves_quantum-mechanical_1987}.
	
The measurement procedure is described by the non-unitary measurement operator,
\begin{eqnarray}
\Upsilon_e(\XM) &=& \Sij{e}{\mathcal{T}\Exp{-\frac{i}{\hbar}\int_0^\tau H_I(t') dt'}}{g},
\end{eqnarray}
where $\mathcal{T}$ denotes temporal ordering. As the operator is diagonal in $\XM$, it is easily solved when resolved in the position basis, $\Upsilon_e(\XM) = \int dx \Upsilon_e(x) \ketbra{x}{x}$. Solving Eqs.~\eqref{eq:ODE} with initial conditions $[c_g,c_e] = [1, 0]$ is equivalent solving the time ordered unitary acting on the initial state $\ket{g}$. The measurement operator is then simply the projection of this state onto $\bra{e}$, giving $\Upsilon_e(x) = c_e(x, \tau)$.

The conditional state of the oscillator following the measurement is $\ket{\psi(\tau)} \sim \Upsilon_e \ket{\psi(0)}$, naturally generalizing to mixed states via $\rho(\tau) \sim \Upsilon_e\rho(0)\Upsilon^\dagger_e$. The normalisation of the final quantum state is simply the probability of realising the measurement operator $\Upsilon_e$, i.e. $\Pr(e) = \mbox{Tr}[\Upsilon_e^\dagger\Upsilon_e\rho_M]$, in accordance with the Born rule~\cite{vanner_pulsed_2011,wiseman_quantum_2009}. The complimentary measurement operator $\Upsilon_g$, corresponds to finding the qubit in the ground state after the measurement. For the present work only $\Upsilon_e$ is of interest and we shall drop the subscript $e$ with the convention that $\Upsilon$ refers to conditioning on the excited state.

\textit{Arbitrary measurement operator} --- To understand how the time dependent drive changes the oscillator's wavefunction, it is helpful to first consider time independent drive, $\alpha(t) = \alpha_0$. In this case the ODE's, Eqs.~\eqref{eq:ODE}, are analytically solvable and result in Rabi oscillations of the qubit with an effective detuning $\Delta_{\makesub{eff}} = {\lambda}_0\XM$. 
Different $\XM$ eigenstates in superposition result in a superposition of effective detunings and Rabi frequencies.

In the following we will restrict the pulse to have unsigned area $\pi/2$. For the constant drive this corresponds a $\pi$-pulse, $\alpha_0 \tau = \pi/2$, with a generalized $\pi$-like pulse requiring $\int dt |{\alpha}(t)| = \pi/2$. At the completion of the $\pi$-pulse, at time $\tau  = \pi/(2{\alpha}_0)$, the measurement operator is given by the Rabi amplitude which may be written as
\begin{equation}
\Upsilon = \frac{\pi}{2}\mathrm{sinc}\left[\frac{\pi}{2}\sqrt{1 + \frac{{\lambda}_0^2\XM^2}{{\alpha}_0^2}}\right] \approx \frac{\pi}{2}\mathrm{sinc}\left[\frac{\pi{\lambda}_0}{2{\alpha}_0}\XM\right].
\label{eq:UpsApprox}
\end{equation}
The sinc function arises via rewriting the Rabi amplitude, $\alpha_0\sin(\Omega t)/\Omega =  \alpha_0t~\mbox{sinc}(\Omega t)$ with Rabi frequency $\Omega^2 = \alpha_0^2 + \lambda_0^2\XM^2 $. Curiously, a sinc function is the Fourier transform of the driving amplitude $\alpha(t)$ (here a top-hat function) with the time-frequency relation clarified by considering $\Upsilon$ a function of ${\lambda}_0 \XM$. With this Fourier transform relation in mind, we postulate the following:

{{The functional form of the measurement operator $\Upsilon(x)$ is well approximated a Fourier transform of the drive amplitude ${\alpha(t)}$}}.

There is no general solution to Eqs.~\eqref{eq:ODE} for arbitrary drive $\alpha(t)$ hence this statement remains a conjecture. Clearly this postulate is not exact as an approximation was made in Eq.~\eqref{eq:UpsApprox} to obtain sinc($x$), however as we shall see, the approximation works remarkably well. It is a very strange relation that the $c_e$ solution of Eqs.~\eqref{eq:ODE} when understood as a function of $x$, gives the Fourier transform of the $\pi$-like pulse. The method only works for the previously given initial conditions, and requires the $\pi$-like pulse normalization. 

For a general pulse shape $\alpha(t)$, the measurement operator may be solved numerically at discrete values of $x$. Unfortunately this numerical solution does not give any technical insight as to why the conjecture seems to work. Generation of any target measurement operator is now realised via Fourier transforming the measurement operator to find the required pulse shape. For technical reason we introduce a dimensionless scaling parameter $\chi$, of order unity, to scale the bandwidth of the drive, $\alpha(t)\rightarrow  \alpha(\chi t)$. The scaling $\chi$ is readily understood in the Fourier transform, changing the characteristic width of the measurement operator (e.g. the factor of $\pi/2$ in Eq.~\eqref{eq:UpsApprox}) and is unique for each target operator. The drive amplitude required to realize a target (superscript $T$) measurement operator $\Upsilon^T(\XM)$ is
\begin{eqnarray}
\alpha(\chi t)  \propto \displaystyle \int_{-\infty}^\infty dx' e^{-i \chi t {\lambda}_0 x'}\Upsilon^T(x') \label{eq:2.alpha}
\end{eqnarray}
with the proportionality constant set to ensure an unsigned pulse area of $\pi/2$. The realised (superscript $R$) measurement operator $\Upsilon^{R}$, is obtained by solving Eqs.~\eqref{eq:ODE} given $\alpha(\chi t)$, and $\chi$ is fixed by maximizing the fidelity between $\Upsilon^{R}$ and $\Upsilon^{T}$. 


\begin{figure}
	\centering
	\includegraphics[width=\columnwidth]{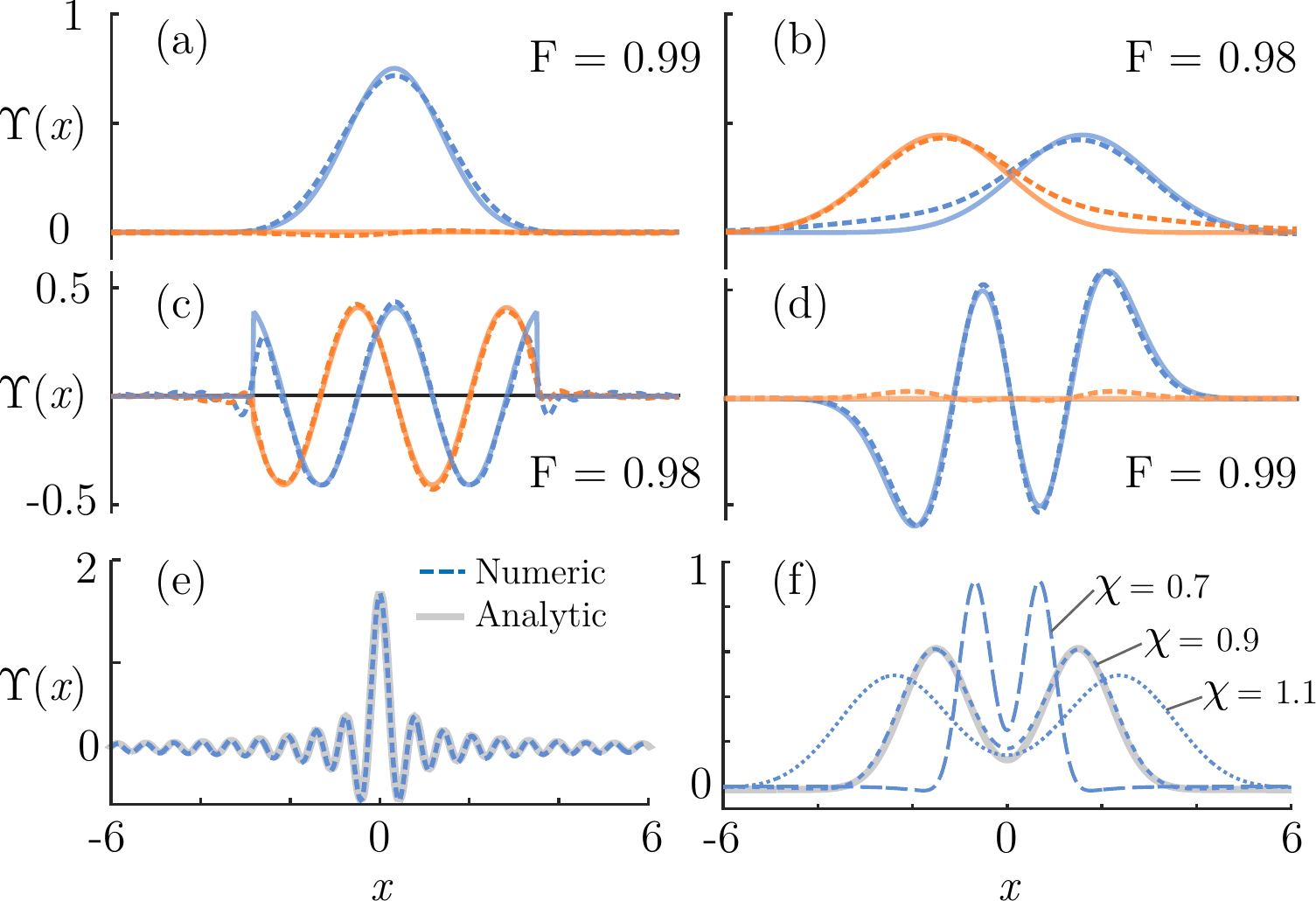}
	\centering
	\caption{Difference between target (solid lines) and numerically obtained realised (dashed lines) measurement operators, separating real (blue) and imaginary (orange) contributions. Target functions are (a) Ground state, (b) Superposition state with relative phase, (c) Plane wave truncated by between $[-3,3]$, (d) $n = 3$ number state. (e) Comparison between the analytic solution from Eq.~\eqref{eq:UpsApprox} and numerical solution for a top-hat drive. (f)  The effect of bandwidth scaling $\chi$ for a two lobed measurement operator, with the target operator plotted in gray. The fidelity is calculated as $|\int dx \Upsilon^T(x) \Upsilon^R(x)^*|$ where each $\Upsilon$ has been normalised to unity using the L2 norm. The ratio of $\lambda_0/\omegam = 0.03$ from Ref.~\cite{khosla_displacemon_2018} is used for all plots.}
	\label{Fig_Target}
\end{figure}

\begin{figure}[hhhhht!!!!!!]
	\centering
	\includegraphics[width=\columnwidth]{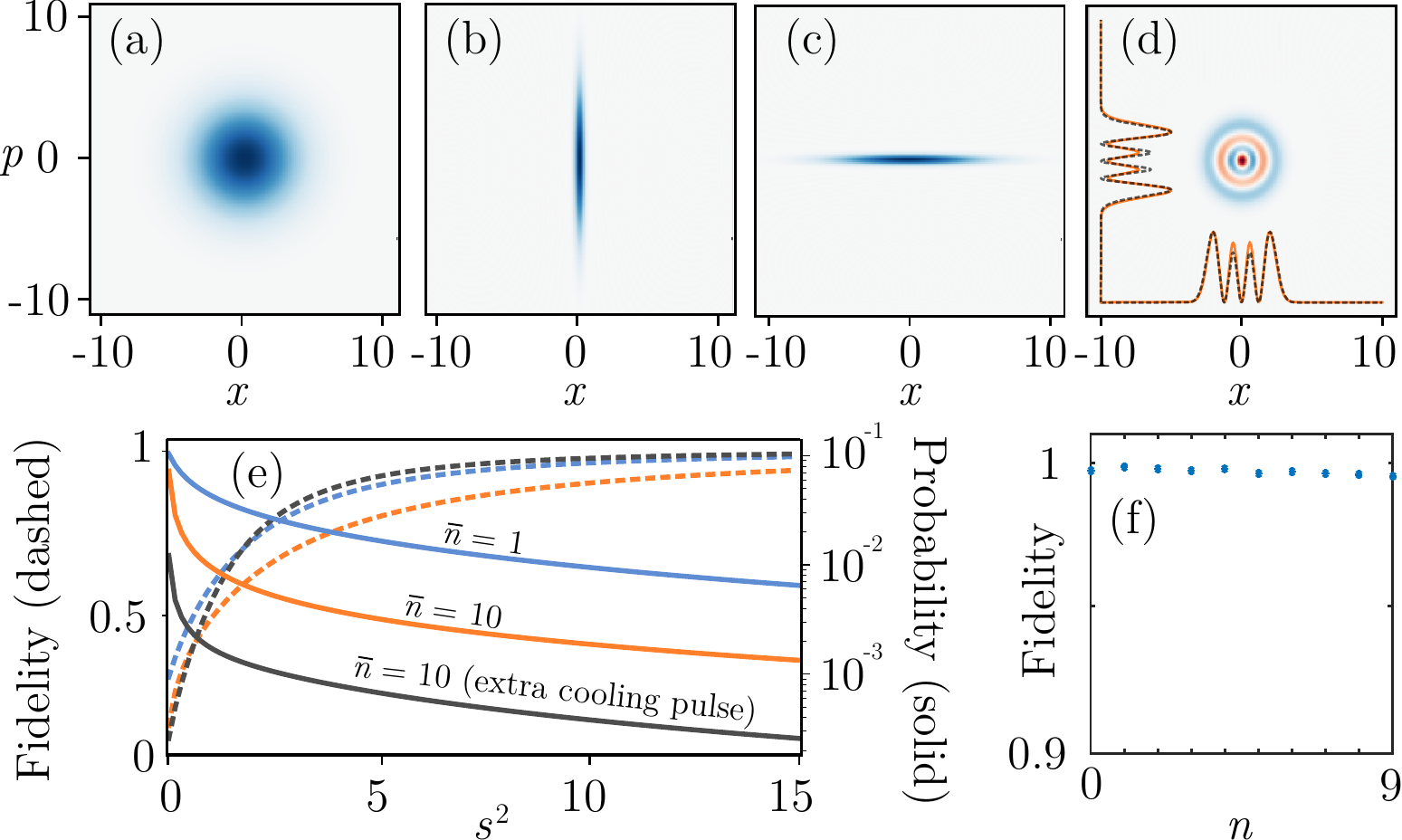}
	\caption{(a-d) Generation of an arbitrary wavefunction from an initial thermal state, plotting Wigner functions at each stage of the process. (a) The initial thermal state with $\nbar = 5$ thermal phonons. (b) A Gaussian measurement is used to position squeeze the oscillator. (c) After one quarter period of evolution, the position squeezing is momentum squeezing, giving a broad Gaussian in the position basis. (d) A second measurement operator is applied to generate the desired quantum state, here an $n = 3$ Fock state. The solid orange curves show the position and momentum marginals of the conditional quantum state, with the dotted gray line showing the marginals for an ideal Fock state. (e) The effect of the squeezing parameter on the fidelity (dashed curves, left axis), and the probability of obtaining  (solid curves, right axis) an $n=3$ number state. The gray curves indicate a three pulse sequence, with two Gaussian pulses cooling orthogonal quadratures, followed by the state preparation pulse. (f) Fidelity of the realized measurement operator for different Fock state projectors $\Upsilon^T =\ketbra{n}{n}$.}
	\label{Fig_Wigners}
\end{figure}

The choice of $\Upsilon^{T}$ is arbitrary, hence any measurement operator (diagonal in the position basis) can be constructed with deterministic parameters. For example the quadratic measurement operator of Ref.~\cite{brawley_nonlinear_2016}, $\Upsilon \propto \exp[-(\bar{x}^2 - \XM^2)^2]$ can be constructed with a deterministically chosen $\bar{x}$. While $\bar{x}$ is deterministic, the protocol is still probabilistic as it requires post-selection on finding the qubit in the excited state. Figure~\ref{Fig_Target} (a-d) plots the fidelity between the (analytically defined) target, and the (numerically obtained) realised measurement operators. For each plot Eq.~\eqref{eq:2.alpha} is used to find the required drive pulse, which is then used in Eqs.~\ref{eq:ODE} to numerically solve for the realised measurement operator $\Upsilon^R$. 

The protocol works for complex measurement operators, Fig.~\ref{Fig_Target}~(b-c), and even for first order discontinuities in the wavefunction, Fig.~\ref{Fig_Target}~(c). The numerical solution exactly matches the analytical solution of Eq.~\eqref{eq:UpsApprox}, Fig.~\ref{Fig_Target}~(e), suggesting negligible numerical error. The effect of the scaling parameter $\chi$ is shown in Fig.~\ref{Fig_Target} (f) for a two lobed measurement operator -- scaling the width of the realized function $\Upsilon^{R}(x)$ -- which is expected given the Fourier relationship between $\chi$ and $x$ from Eq.~\eqref{eq:2.alpha}.

In this section we have shown how to generate an arbitrary shaped position basis measurement operator, but have not discussed the probabilities of obtaining a successful outcome, or the purity of the final state. Furthermore, while the $\Upsilon$'s shown in Fig.~\ref{Fig_Target} are written as wavefunctions, the measurement operator is diagonal in $\XM$ and hence cannot be understood as an arbitrary projector~\cite{wiseman_quantum_1996}. In the following section we will address these points and show under what conditions $\Upsilon$ results in a target quantum state. 

\textit{Arbitrary quantum states} --- 
%
For generating an arbitrary quantum state we consider a two step preparation procedure (see Fig.~\ref{Fig_Wigners}) similar to Ref.~\cite{vanner_pulsed_2011}. The initial state is assumed to be a thermal state with occupation $\nbar$. The first step applies a Gaussian measurement operator $\Upsilon_1 \propto \exp(-\XM^2s^2/4)$, squeezing the position quadrature (for $s > 1$). This first step constrains the state of the oscillator, reducing it's entropy~\cite{vanner_cooling-by-measurement_2013,khosla_displacemon_2018}. After a quarter period of free mechanical evolution, the state is squeezed in the momentum quadrature, $\rho_s \sim \mathcal{R}(\pi/4)\Upsilon_1 \rho_0 \Upsilon_1^\dagger \mathcal{R}^\dagger(\pi/4)$, where $\mathcal{R}$ is the rotation operator. 

This momentum squeezed state is taken to be the initial state for the second step, where a second measurement operator is chosen to be proportional to the desired wavefunction $\Upsilon_2(x) \propto \psi(x)$. The final state after conditioning on both measurements is $\rho(\tau) \sim \Upsilon_2\rho_s\Upsilon_2^\dagger$. Not only does this second measurement operator give the desired position basis wavefunction, it also further purifies the state. The fidelity between the desired and conditional quantum states is given by
\begin{eqnarray}
F^2 &=&  \frac{\int dx'dx'' \psi(x')\psi^*(x'')  \Sij{x'}{\Upsilon_2\rho_s\Upsilon_2^\dagger}{x''}}{\int dx'  \Sij{x'}{\Upsilon_2\rho_s\Upsilon_2^\dagger}{x'}},
\label{eq:2.FidAct}
\end{eqnarray}
where we may approximate $\Upsilon_2(x) \propto \psi(x')$ up to $\approx 0.98$ fidelity (see Fig.~\ref{Fig_Target} and Fig.~\ref{Fig_Wigners} (f), and note the normalization of $\Upsilon_2$ factors out of $F^2$). In the case of strong squeezing $s^2> \nbar$, the matrix elements 
define a broad Gaussian, centered at $x' = x'' = 0$~\cite{walls_quantum_2008}. If the variance of this Gaussian is large compared to the spatial extent of the wavefunction, then $\Sij{x'}{\rho_s}{x''} \approx \Sij{0}{\rho_s}{0}$ remains approximately constant in the integral over the wavefunction. Under this approximation, the state dependence factors out of the fidelity,
\begin{eqnarray}
F^2 \approx  \frac{\int dx'|\psi(x')|^2  \int dx'' |\psi(x'')|^2}{\int dx' |\psi(x')|^2} \frac{\Sij{0}{\rho_s}{0}}{\Sij{0}{\rho_s}{0}} = 1 \label{eq:2.fidapp}
\end{eqnarray}  
and the target wavefunction can be achieved with near unity Fidelity.  Fig.~\ref{Fig_Wigners} (e) shows the expected fidelity (left axis), taking into account both the Gaussian nature of the initial state and the sub-unity fidelity between the target and realized wavefunctions. As the squeezing parameter increases beyond the initial thermal occupation, the near unit fidelity predicted by Eq.~\eqref{eq:2.fidapp} is observed.

Increasing the squeezing parameter increases the fidelity at the cost of lowering the probability of obtaining the desired measurement operator $\Upsilon$ (as opposed to the complementary operator $\Upsilon_g$). The increased fidelity is a consequence of a broader Gaussian in the position basis, and therefore the curvature of the Gaussian becoming less significant over the width of the wavefunction. However, the probability of obtaining $\Upsilon$ is $\mathrm{Tr}[\Upsilon^\dagger\Upsilon \rho(0)] $, which is exactly the overlap between $|\Upsilon(x)|^2$, and the position basis probability distribution. The broader Gaussian results is a lower overlap integral, and thus reduces the probability of success, plotted on the right axis of Fig~\ref{Fig_Wigners}~(e).

As previously noted, the measurement operator is diagonal in the position basis, hence this form of state preparation should be understood as wavefunction shaping as opposed to von Neumann measurement. For example it cannot be understood as a Fock state projector $\ketbra{n}{n}$, but rather constrains the position basis probability to be that of a number state. As a pure state is uniquely identified by it's wavefunction this implies the resulting state must be a Fock state.

\textit{Non-unitarity} --- Non-unitary effects such as qubit dephasing and mechanical thermalization have been neglected without sufficient justification. In the following we show that the entire protocol may be done within the coherence time of the qubit-oscillator system. We take the experimentally accessible parameters given in Ref.~\cite{khosla_displacemon_2018}:  $\omegam/2\pi = 125$~MHz, $\lambda_0/2\pi = 8.5$~MHz, qubit dephasing time $T_2 = 2~\mu$s, and mechanical quality factor $Q = 10^5$, with the system in equilibrium at 33 mK, $\nbar = 5$. 

For these parameters the mechanical coherence time is $100~\mu$s, hence the system is limited by the qubit's $T_2$ lifetime. If the pulse duration is less than 200~ns, the system is well approximated as unitary. Note the pulse is a Fourier transform of the desired wavefunction, hence the reciprocal scaling of the Fourier transform can be used to estimate the pulse duration for any target wavefunction. Ground state (or number state) conditioning require features in the spectrum of $\XM$ on the order of unity, which in reciprocal spare requires $\chi\lambda_0 t \approx 1$ (see Eq.~\eqref{eq:2.alpha} and recall $\chi$ is of order unity). Hence the characteristic width of the pulse is $2\pi\lambda_0^{-1} \approx 11$~ns, well within the coherence time of the qubit. This also gives the general condition for unitarity, $\lambda_0 \gg 2\pi/T_2, \nbar\omegam/Q$, which is simply the condition for strong coupling. 

When the coherence time is constrained by the qubit's lifetime, several coherent measurement operators may be realized within the resonator's coherence time. For example two Gaussian measurement operators of width $s^{-2}$ can be applied resulting in an effective single Gaussian operator of width $s^{-2}/\sqrt{2}$. The scaling is unfavorable for repeated Gaussian pulses, however each measurement may be applied to orthogonal quadratures enabling fast cooling~\cite{vanner_cooling-by-measurement_2013}. The gray curves in Fig.~\ref{Fig_Wigners}(e) use this two step orthogonal quadrature cooling before the state conditioning measurement. This results in a significantly larger fidelity, attributable to better state purification, however this comes at the price of lowering success probability. 

Resolving sub-ground state features, such as projecting onto large momentum cat states~\cite{vanner_selective_2011}, requires proportionally longer pulse durations. However, note that momentum and position quadrature wavefunctions form Fourier transform pairs. With this in mind, one can generate sub-ground state features in the momentum quadrature requiring only ground state features in the position quadrature (e.g. see the position cat state in Fig.~\ref{Fig_Target}(b) -- cat states have sub-ground state features in their wavefunction~\cite{brawley_nonlinear_2016,clarke_growing_2018}). In this way, any desired wavefunction can be conditioned to the extent that it has support in the initial thermal state.

\textit{Summary} --- We have introduced a method to generate an arbitrary position basis measurement operator in a coupled qubit-mechanical system. Although the equations of motion do not have an analytic solution, we have shown that a numerical procedure can generate the drive pulse required to realise any target measurement operator with $\gtrsim 0.98$ fidelity. The protocol requires parameters well into the strong coupling regime, and can be achieved in principle by combining state of the art qubits~\cite{braumuller_concentric_2016,mallet_single-shot_2009} and resonators~\cite{moser_nanotube_2014}. Realizing a general position basis measurement operator enables full quantum state tomography via standard techniques~\cite{vogel_determination_1989,smithey_measurement_1993,lvovsky_continuous-variable_2009}.

Using a two step measurement procedure, we showed how to construct an arbitrary quantum state with $\gtrsim 0.95$ fidelity from an initial thermal state. The first step required partial purification of the Gaussian state, and the second involved projecting the desired wavefuncion, enabling generation of an arbitrary wavefunction of the mechanical oscillator. Such a scheme naturally generalizes to a multi-step purification/generation protocol. The general technique relies on the non-linear Hamiltonian negating the requirement for quantum control of the microwave field. The prepared quantum state can be used as a resource~\cite{yadin_operational_2018} which can be swapped into the microwave field~\cite{reed_faithful_2017} and used for quantum information tasks.

\subsection{Acknowledgments}

I'd like to thank Edward A. Laird and Gerard J. Milburn for helpful discussions, and in particular E. Laird for his key insights in motivating this work. Thanks also to M. R. Vanner, C. Wood and A. L. Khosla for constructive comments.

\bibliography{bib_keep}

\begin{thebibliography}{61}%
\makeatletter
\providecommand \@ifxundefined [1]{%
 \@ifx{#1\undefined}
}%
\providecommand \@ifnum [1]{%
 \ifnum #1\expandafter \@firstoftwo
 \else \expandafter \@secondoftwo
 \fi
}%
\providecommand \@ifx [1]{%
 \ifx #1\expandafter \@firstoftwo
 \else \expandafter \@secondoftwo
 \fi
}%
\providecommand \natexlab [1]{#1}%
\providecommand \enquote  [1]{``#1''}%
\providecommand \bibnamefont  [1]{#1}%
\providecommand \bibfnamefont [1]{#1}%
\providecommand \citenamefont [1]{#1}%
\providecommand \href@noop [0]{\@secondoftwo}%
\providecommand \href [0]{\begingroup \@sanitize@url \@href}%
\providecommand \@href[1]{\@@startlink{#1}\@@href}%
\providecommand \@@href[1]{\endgroup#1\@@endlink}%
\providecommand \@sanitize@url [0]{\catcode `\\12\catcode `\$12\catcode
  `\&12\catcode `\#12\catcode `\^12\catcode `\_12\catcode `\%12\relax}%
\providecommand \@@startlink[1]{}%
\providecommand \@@endlink[0]{}%
\providecommand \url  [0]{\begingroup\@sanitize@url \@url }%
\providecommand \@url [1]{\endgroup\@href {#1}{\urlprefix }}%
\providecommand \urlprefix  [0]{URL }%
\providecommand \Eprint [0]{\href }%
\providecommand \doibase [0]{http://dx.doi.org/}%
\providecommand \selectlanguage [0]{\@gobble}%
\providecommand \bibinfo  [0]{\@secondoftwo}%
\providecommand \bibfield  [0]{\@secondoftwo}%
\providecommand \translation [1]{[#1]}%
\providecommand \BibitemOpen [0]{}%
\providecommand \bibitemStop [0]{}%
\providecommand \bibitemNoStop [0]{.\EOS\space}%
\providecommand \EOS [0]{\spacefactor3000\relax}%
\providecommand \BibitemShut  [1]{\csname bibitem#1\endcsname}%
\let\auto@bib@innerbib\@empty
\bibitem [{\citenamefont {Aspelmeyer}\ \emph {et~al.}(2014)\citenamefont
  {Aspelmeyer}, \citenamefont {Kippenberg},\ and\ \citenamefont
  {Marquardt}}]{aspelmeyer_cavity_2014}%
  \BibitemOpen
  \bibfield  {author} {\bibinfo {author} {\bibfnamefont {M.}~\bibnamefont
  {Aspelmeyer}}, \bibinfo {author} {\bibfnamefont {T.~J.}\ \bibnamefont
  {Kippenberg}}, \ and\ \bibinfo {author} {\bibfnamefont {F.}~\bibnamefont
  {Marquardt}},\ }\href {\doibase 10.1103/RevModPhys.86.1391} {\bibfield
  {journal} {\bibinfo  {journal} {Rev. Mod. Phys.}\ }\textbf {\bibinfo {volume}
  {86}},\ \bibinfo {pages} {1391} (\bibinfo {year} {2014})}\BibitemShut
  {NoStop}%
\bibitem [{\citenamefont {Pirkkalainen}\ \emph {et~al.}(2013)\citenamefont
  {Pirkkalainen}, \citenamefont {Cho}, \citenamefont {Li}, \citenamefont
  {Paraoanu}, \citenamefont {Hakonen},\ and\ \citenamefont
  {Sillanpaa}}]{pirkkalainen_hybrid_2013}%
  \BibitemOpen
  \bibfield  {author} {\bibinfo {author} {\bibfnamefont {J.-M.}\ \bibnamefont
  {Pirkkalainen}}, \bibinfo {author} {\bibfnamefont {S.~U.}\ \bibnamefont
  {Cho}}, \bibinfo {author} {\bibfnamefont {J.}~\bibnamefont {Li}}, \bibinfo
  {author} {\bibfnamefont {G.~S.}\ \bibnamefont {Paraoanu}}, \bibinfo {author}
  {\bibfnamefont {P.~J.}\ \bibnamefont {Hakonen}}, \ and\ \bibinfo {author}
  {\bibfnamefont {M.~A.}\ \bibnamefont {Sillanpaa}},\ }\href {\doibase
  10.1038/nature11821} {\bibfield  {journal} {\bibinfo  {journal} {Nature}\
  }\textbf {\bibinfo {volume} {494}},\ \bibinfo {pages} {211} (\bibinfo {year}
  {2013})}\BibitemShut {NoStop}%
\bibitem [{\citenamefont {Lecocq}\ \emph {et~al.}(2015)\citenamefont {Lecocq},
  \citenamefont {Teufel}, \citenamefont {Aumentado},\ and\ \citenamefont
  {Simmonds}}]{lecocq_resolving_2015}%
  \BibitemOpen
  \bibfield  {author} {\bibinfo {author} {\bibfnamefont {F.}~\bibnamefont
  {Lecocq}}, \bibinfo {author} {\bibfnamefont {J.~D.}\ \bibnamefont {Teufel}},
  \bibinfo {author} {\bibfnamefont {J.}~\bibnamefont {Aumentado}}, \ and\
  \bibinfo {author} {\bibfnamefont {R.~W.}\ \bibnamefont {Simmonds}},\ }\href
  {\doibase 10.1038/nphys3365} {\bibfield  {journal} {\bibinfo  {journal} {Nat
  Phys}\ }\textbf {\bibinfo {volume} {11}},\ \bibinfo {pages} {635} (\bibinfo
  {year} {2015})}\BibitemShut {NoStop}%
\bibitem [{\citenamefont {Pirkkalainen}\ \emph {et~al.}(2015)\citenamefont
  {Pirkkalainen}, \citenamefont {Cho}, \citenamefont {Massel}, \citenamefont
  {Tuorila}, \citenamefont {Heikkila}, \citenamefont {Hakonen},\ and\
  \citenamefont {Sillanpaa}}]{pirkkalainen_cavity_2015}%
  \BibitemOpen
  \bibfield  {author} {\bibinfo {author} {\bibfnamefont {J.-M.}\ \bibnamefont
  {Pirkkalainen}}, \bibinfo {author} {\bibfnamefont {S.}~\bibnamefont {Cho}},
  \bibinfo {author} {\bibfnamefont {F.}~\bibnamefont {Massel}}, \bibinfo
  {author} {\bibfnamefont {J.}~\bibnamefont {Tuorila}}, \bibinfo {author}
  {\bibfnamefont {T.}~\bibnamefont {Heikkila}}, \bibinfo {author}
  {\bibfnamefont {P.}~\bibnamefont {Hakonen}}, \ and\ \bibinfo {author}
  {\bibfnamefont {M.}~\bibnamefont {Sillanpaa}},\ }\href {\doibase
  10.1038/ncomms7981} {\bibfield  {journal} {\bibinfo  {journal} {Nat Commun}\
  }\textbf {\bibinfo {volume} {6}} (\bibinfo {year} {2015}),\
  10.1038/ncomms7981}\BibitemShut {NoStop}%
\bibitem [{\citenamefont {Heikkila}\ \emph {et~al.}(2014)\citenamefont
  {Heikkila}, \citenamefont {Massel}, \citenamefont {Tuorila}, \citenamefont
  {Khan},\ and\ \citenamefont {Sillanpaa}}]{heikkila_enhancing_2014}%
  \BibitemOpen
  \bibfield  {author} {\bibinfo {author} {\bibfnamefont {T.~T.}\ \bibnamefont
  {Heikkila}}, \bibinfo {author} {\bibfnamefont {F.}~\bibnamefont {Massel}},
  \bibinfo {author} {\bibfnamefont {J.}~\bibnamefont {Tuorila}}, \bibinfo
  {author} {\bibfnamefont {R.}~\bibnamefont {Khan}}, \ and\ \bibinfo {author}
  {\bibfnamefont {M.~A.}\ \bibnamefont {Sillanpaa}},\ }\href {\doibase
  10.1103/PhysRevLett.112.203603} {\bibfield  {journal} {\bibinfo  {journal}
  {Physical Review Letters}\ }\textbf {\bibinfo {volume} {112}} (\bibinfo
  {year} {2014}),\ 10.1103/PhysRevLett.112.203603},\ \bibinfo {note} {arXiv:
  1311.3802}\BibitemShut {NoStop}%
\bibitem [{\citenamefont {Sillanpaa}\ \emph {et~al.}(2009)\citenamefont
  {Sillanpaa}, \citenamefont {Sarkar}, \citenamefont {Sulkko}, \citenamefont
  {Muhonen},\ and\ \citenamefont {Hakonen}}]{sillanpaa_accessing_2009}%
  \BibitemOpen
  \bibfield  {author} {\bibinfo {author} {\bibfnamefont {M.~A.}\ \bibnamefont
  {Sillanpaa}}, \bibinfo {author} {\bibfnamefont {J.}~\bibnamefont {Sarkar}},
  \bibinfo {author} {\bibfnamefont {J.}~\bibnamefont {Sulkko}}, \bibinfo
  {author} {\bibfnamefont {J.}~\bibnamefont {Muhonen}}, \ and\ \bibinfo
  {author} {\bibfnamefont {P.~J.}\ \bibnamefont {Hakonen}},\ }\href {\doibase
  10.1063/1.3173826} {\bibfield  {journal} {\bibinfo  {journal} {Appl. Phys.
  Lett.}\ }\textbf {\bibinfo {volume} {95}},\ \bibinfo {pages} {011909}
  (\bibinfo {year} {2009})}\BibitemShut {NoStop}%
\bibitem [{\citenamefont {Teufel}\ \emph {et~al.}(2011)\citenamefont {Teufel},
  \citenamefont {Li}, \citenamefont {Allman}, \citenamefont {Cicak},
  \citenamefont {Sirois}, \citenamefont {Whittaker},\ and\ \citenamefont
  {Simmonds}}]{teufel_circuit_2011}%
  \BibitemOpen
  \bibfield  {author} {\bibinfo {author} {\bibfnamefont {J.~D.}\ \bibnamefont
  {Teufel}}, \bibinfo {author} {\bibfnamefont {D.}~\bibnamefont {Li}}, \bibinfo
  {author} {\bibfnamefont {M.~S.}\ \bibnamefont {Allman}}, \bibinfo {author}
  {\bibfnamefont {K.}~\bibnamefont {Cicak}}, \bibinfo {author} {\bibfnamefont
  {A.~J.}\ \bibnamefont {Sirois}}, \bibinfo {author} {\bibfnamefont {J.~D.}\
  \bibnamefont {Whittaker}}, \ and\ \bibinfo {author} {\bibfnamefont {R.~W.}\
  \bibnamefont {Simmonds}},\ }\href {\doibase 10.1038/nature09898} {\bibfield
  {journal} {\bibinfo  {journal} {Nature}\ }\textbf {\bibinfo {volume} {471}},\
  \bibinfo {pages} {204} (\bibinfo {year} {2011})}\BibitemShut {NoStop}%
\bibitem [{\citenamefont {Armour}\ and\ \citenamefont
  {Blencowe}(2008)}]{armour_probing_2008}%
  \BibitemOpen
  \bibfield  {author} {\bibinfo {author} {\bibfnamefont {A.~D.}\ \bibnamefont
  {Armour}}\ and\ \bibinfo {author} {\bibfnamefont {M.~P.}\ \bibnamefont
  {Blencowe}},\ }\href {\doibase 10.1088/1367-2630/10/9/095004} {\bibfield
  {journal} {\bibinfo  {journal} {New J. Phys.}\ }\textbf {\bibinfo {volume}
  {10}},\ \bibinfo {pages} {095004} (\bibinfo {year} {2008})}\BibitemShut
  {NoStop}%
\bibitem [{\citenamefont {Armour}\ \emph {et~al.}(2002)\citenamefont {Armour},
  \citenamefont {Blencowe},\ and\ \citenamefont
  {Schwab}}]{armour_entanglement_2002}%
  \BibitemOpen
  \bibfield  {author} {\bibinfo {author} {\bibfnamefont {A.~D.}\ \bibnamefont
  {Armour}}, \bibinfo {author} {\bibfnamefont {M.~P.}\ \bibnamefont
  {Blencowe}}, \ and\ \bibinfo {author} {\bibfnamefont {K.~C.}\ \bibnamefont
  {Schwab}},\ }\href {\doibase 10.1103/PhysRevLett.88.148301} {\bibfield
  {journal} {\bibinfo  {journal} {Phys. Rev. Lett.}\ }\textbf {\bibinfo
  {volume} {88}},\ \bibinfo {pages} {148301} (\bibinfo {year}
  {2002})}\BibitemShut {NoStop}%
\bibitem [{\citenamefont {Blencowe}\ and\ \citenamefont
  {Armour}(2008)}]{blencowe_probing_2008}%
  \BibitemOpen
  \bibfield  {author} {\bibinfo {author} {\bibfnamefont {M.~P.}\ \bibnamefont
  {Blencowe}}\ and\ \bibinfo {author} {\bibfnamefont {A.~D.}\ \bibnamefont
  {Armour}},\ }\href {\doibase 10.1088/1367-2630/10/9/095005} {\bibfield
  {journal} {\bibinfo  {journal} {New J. Phys.}\ }\textbf {\bibinfo {volume}
  {10}},\ \bibinfo {pages} {095005} (\bibinfo {year} {2008})}\BibitemShut
  {NoStop}%
\bibitem [{\citenamefont
  {Blencowe}(2005)}]{blencowe_nanoelectromechanical_2005}%
  \BibitemOpen
  \bibfield  {author} {\bibinfo {author} {\bibfnamefont {M.~P.}\ \bibnamefont
  {Blencowe}},\ }\href {\doibase 10.1080/00107510500146865} {\bibfield
  {journal} {\bibinfo  {journal} {Contemporary Physics}\ }\textbf {\bibinfo
  {volume} {46}},\ \bibinfo {pages} {249} (\bibinfo {year} {2005})}\BibitemShut
  {NoStop}%
\bibitem [{\citenamefont {Safavi-Naeini}\ \emph {et~al.}(2012)\citenamefont
  {Safavi-Naeini}, \citenamefont {Chan}, \citenamefont {Hill}, \citenamefont
  {Alegre}, \citenamefont {Krause},\ and\ \citenamefont
  {Painter}}]{safavi-naeini_observation_2012}%
  \BibitemOpen
  \bibfield  {author} {\bibinfo {author} {\bibfnamefont {A.~H.}\ \bibnamefont
  {Safavi-Naeini}}, \bibinfo {author} {\bibfnamefont {J.}~\bibnamefont {Chan}},
  \bibinfo {author} {\bibfnamefont {J.~T.}\ \bibnamefont {Hill}}, \bibinfo
  {author} {\bibfnamefont {T.~P.~M.}\ \bibnamefont {Alegre}}, \bibinfo {author}
  {\bibfnamefont {A.}~\bibnamefont {Krause}}, \ and\ \bibinfo {author}
  {\bibfnamefont {O.}~\bibnamefont {Painter}},\ }\href {\doibase
  10.1103/PhysRevLett.108.033602} {\bibfield  {journal} {\bibinfo  {journal}
  {Phys. Rev. Lett.}\ }\textbf {\bibinfo {volume} {108}},\ \bibinfo {pages}
  {033602} (\bibinfo {year} {2012})}\BibitemShut {NoStop}%
\bibitem [{\citenamefont {O`Connell}\ \emph {et~al.}(2010)\citenamefont
  {O`Connell}, \citenamefont {Hofheinz}, \citenamefont {Ansmann}, \citenamefont
  {Bialczak}, \citenamefont {Lenander}, \citenamefont {Lucero}, \citenamefont
  {Neeley}, \citenamefont {Sank}, \citenamefont {Wang}, \citenamefont {Weides},
  \citenamefont {Wenner}, \citenamefont {Martinis},\ and\ \citenamefont
  {Cleland}}]{oconnell_quantum_2010}%
  \BibitemOpen
  \bibfield  {author} {\bibinfo {author} {\bibfnamefont {A.~D.}\ \bibnamefont
  {O`Connell}}, \bibinfo {author} {\bibfnamefont {M.}~\bibnamefont {Hofheinz}},
  \bibinfo {author} {\bibfnamefont {M.}~\bibnamefont {Ansmann}}, \bibinfo
  {author} {\bibfnamefont {R.~C.}\ \bibnamefont {Bialczak}}, \bibinfo {author}
  {\bibfnamefont {M.}~\bibnamefont {Lenander}}, \bibinfo {author}
  {\bibfnamefont {E.}~\bibnamefont {Lucero}}, \bibinfo {author} {\bibfnamefont
  {M.}~\bibnamefont {Neeley}}, \bibinfo {author} {\bibfnamefont
  {D.}~\bibnamefont {Sank}}, \bibinfo {author} {\bibfnamefont {H.}~\bibnamefont
  {Wang}}, \bibinfo {author} {\bibfnamefont {M.}~\bibnamefont {Weides}},
  \bibinfo {author} {\bibfnamefont {J.}~\bibnamefont {Wenner}}, \bibinfo
  {author} {\bibfnamefont {J.~M.}\ \bibnamefont {Martinis}}, \ and\ \bibinfo
  {author} {\bibfnamefont {A.~N.}\ \bibnamefont {Cleland}},\ }\href {\doibase
  10.1038/nature08967} {\bibfield  {journal} {\bibinfo  {journal} {Nature}\
  }\textbf {\bibinfo {volume} {464}},\ \bibinfo {pages} {697} (\bibinfo {year}
  {2010})}\BibitemShut {NoStop}%
\bibitem [{\citenamefont {Wollman}\ \emph {et~al.}(2015)\citenamefont
  {Wollman}, \citenamefont {Lei}, \citenamefont {Weinstein}, \citenamefont
  {Suh}, \citenamefont {Kronwald}, \citenamefont {Marquardt}, \citenamefont
  {Clerk},\ and\ \citenamefont {Schwab}}]{wollman_quantum_2015}%
  \BibitemOpen
  \bibfield  {author} {\bibinfo {author} {\bibfnamefont {E.~E.}\ \bibnamefont
  {Wollman}}, \bibinfo {author} {\bibfnamefont {C.~U.}\ \bibnamefont {Lei}},
  \bibinfo {author} {\bibfnamefont {A.~J.}\ \bibnamefont {Weinstein}}, \bibinfo
  {author} {\bibfnamefont {J.}~\bibnamefont {Suh}}, \bibinfo {author}
  {\bibfnamefont {A.}~\bibnamefont {Kronwald}}, \bibinfo {author}
  {\bibfnamefont {F.}~\bibnamefont {Marquardt}}, \bibinfo {author}
  {\bibfnamefont {A.~A.}\ \bibnamefont {Clerk}}, \ and\ \bibinfo {author}
  {\bibfnamefont {K.~C.}\ \bibnamefont {Schwab}},\ }\href {\doibase
  10.1126/science.aac5138} {\bibfield  {journal} {\bibinfo  {journal}
  {Science}\ }\textbf {\bibinfo {volume} {349}},\ \bibinfo {pages} {952}
  (\bibinfo {year} {2015})}\BibitemShut {NoStop}%
\bibitem [{\citenamefont {Palomaki}\ \emph
  {et~al.}(2013{\natexlab{a}})\citenamefont {Palomaki}, \citenamefont {Teufel},
  \citenamefont {Simmonds},\ and\ \citenamefont
  {Lehnert}}]{palomaki_entangling_2013}%
  \BibitemOpen
  \bibfield  {author} {\bibinfo {author} {\bibfnamefont {T.~A.}\ \bibnamefont
  {Palomaki}}, \bibinfo {author} {\bibfnamefont {J.~D.}\ \bibnamefont
  {Teufel}}, \bibinfo {author} {\bibfnamefont {R.~W.}\ \bibnamefont
  {Simmonds}}, \ and\ \bibinfo {author} {\bibfnamefont {K.~W.}\ \bibnamefont
  {Lehnert}},\ }\href {\doibase 10.1126/science.1244563} {\bibfield  {journal}
  {\bibinfo  {journal} {Science}\ ,\ \bibinfo {pages} {1244563}} (\bibinfo
  {year} {2013}{\natexlab{a}})}\BibitemShut {NoStop}%
\bibitem [{\citenamefont {Riedinger}\ \emph {et~al.}(2016)\citenamefont
  {Riedinger}, \citenamefont {Hong}, \citenamefont {Norte}, \citenamefont
  {Slater}, \citenamefont {Shang}, \citenamefont {Krause}, \citenamefont
  {Anant}, \citenamefont {Aspelmeyer},\ and\ \citenamefont
  {Groblacher}}]{riedinger_non-classical_2016}%
  \BibitemOpen
  \bibfield  {author} {\bibinfo {author} {\bibfnamefont {R.}~\bibnamefont
  {Riedinger}}, \bibinfo {author} {\bibfnamefont {S.}~\bibnamefont {Hong}},
  \bibinfo {author} {\bibfnamefont {R.~A.}\ \bibnamefont {Norte}}, \bibinfo
  {author} {\bibfnamefont {J.~A.}\ \bibnamefont {Slater}}, \bibinfo {author}
  {\bibfnamefont {J.}~\bibnamefont {Shang}}, \bibinfo {author} {\bibfnamefont
  {A.~G.}\ \bibnamefont {Krause}}, \bibinfo {author} {\bibfnamefont
  {V.}~\bibnamefont {Anant}}, \bibinfo {author} {\bibfnamefont
  {M.}~\bibnamefont {Aspelmeyer}}, \ and\ \bibinfo {author} {\bibfnamefont
  {S.}~\bibnamefont {Groblacher}},\ }\href {\doibase 10.1038/nature16536}
  {\bibfield  {journal} {\bibinfo  {journal} {Nature}\ }\textbf {\bibinfo
  {volume} {530}},\ \bibinfo {pages} {313} (\bibinfo {year}
  {2016})}\BibitemShut {NoStop}%
\bibitem [{\citenamefont {Kerckhoff}\ \emph {et~al.}(2013)\citenamefont
  {Kerckhoff}, \citenamefont {Andrews}, \citenamefont {Ku}, \citenamefont
  {Kindel}, \citenamefont {Cicak}, \citenamefont {Simmonds},\ and\
  \citenamefont {Lehnert}}]{kerckhoff_tunable_2013}%
  \BibitemOpen
  \bibfield  {author} {\bibinfo {author} {\bibfnamefont {J.}~\bibnamefont
  {Kerckhoff}}, \bibinfo {author} {\bibfnamefont {R.~W.}\ \bibnamefont
  {Andrews}}, \bibinfo {author} {\bibfnamefont {H.~S.}\ \bibnamefont {Ku}},
  \bibinfo {author} {\bibfnamefont {W.~F.}\ \bibnamefont {Kindel}}, \bibinfo
  {author} {\bibfnamefont {K.}~\bibnamefont {Cicak}}, \bibinfo {author}
  {\bibfnamefont {R.~W.}\ \bibnamefont {Simmonds}}, \ and\ \bibinfo {author}
  {\bibfnamefont {K.~W.}\ \bibnamefont {Lehnert}},\ }\href {\doibase
  10.1103/PhysRevX.3.021013} {\bibfield  {journal} {\bibinfo  {journal} {Phys.
  Rev. X}\ }\textbf {\bibinfo {volume} {3}},\ \bibinfo {pages} {021013}
  (\bibinfo {year} {2013})}\BibitemShut {NoStop}%
\bibitem [{\citenamefont {Lee}\ \emph {et~al.}(2010)\citenamefont {Lee},
  \citenamefont {McRae}, \citenamefont {Harris}, \citenamefont {Knittel},\ and\
  \citenamefont {Bowen}}]{lee_cooling_2010}%
  \BibitemOpen
  \bibfield  {author} {\bibinfo {author} {\bibfnamefont {K.~H.}\ \bibnamefont
  {Lee}}, \bibinfo {author} {\bibfnamefont {T.~G.}\ \bibnamefont {McRae}},
  \bibinfo {author} {\bibfnamefont {G.~I.}\ \bibnamefont {Harris}}, \bibinfo
  {author} {\bibfnamefont {J.}~\bibnamefont {Knittel}}, \ and\ \bibinfo
  {author} {\bibfnamefont {W.~P.}\ \bibnamefont {Bowen}},\ }\href {\doibase
  10.1103/PhysRevLett.104.123604} {\bibfield  {journal} {\bibinfo  {journal}
  {Phys. Rev. Lett.}\ }\textbf {\bibinfo {volume} {104}},\ \bibinfo {pages}
  {123604} (\bibinfo {year} {2010})}\BibitemShut {NoStop}%
\bibitem [{\citenamefont {Asadian}\ \emph {et~al.}(2014)\citenamefont
  {Asadian}, \citenamefont {Brukner},\ and\ \citenamefont
  {Rabl}}]{asadian_probing_2014}%
  \BibitemOpen
  \bibfield  {author} {\bibinfo {author} {\bibfnamefont {A.}~\bibnamefont
  {Asadian}}, \bibinfo {author} {\bibfnamefont {C.}~\bibnamefont {Brukner}}, \
  and\ \bibinfo {author} {\bibfnamefont {P.}~\bibnamefont {Rabl}},\ }\href
  {\doibase 10.1103/PhysRevLett.112.190402} {\bibfield  {journal} {\bibinfo
  {journal} {Physical Review Letters}\ }\textbf {\bibinfo {volume} {112}}
  (\bibinfo {year} {2014}),\ 10.1103/PhysRevLett.112.190402},\ \bibinfo {note}
  {arXiv: 1309.2229}\BibitemShut {NoStop}%
\bibitem [{\citenamefont {Khosla}\ \emph {et~al.}(2018)\citenamefont {Khosla},
  \citenamefont {Vanner}, \citenamefont {Ares},\ and\ \citenamefont
  {Laird}}]{khosla_displacemon_2018}%
  \BibitemOpen
  \bibfield  {author} {\bibinfo {author} {\bibfnamefont {K.}~\bibnamefont
  {Khosla}}, \bibinfo {author} {\bibfnamefont {M.}~\bibnamefont {Vanner}},
  \bibinfo {author} {\bibfnamefont {N.}~\bibnamefont {Ares}}, \ and\ \bibinfo
  {author} {\bibfnamefont {E.}~\bibnamefont {Laird}},\ }\href {\doibase
  10.1103/PhysRevX.8.021052} {\bibfield  {journal} {\bibinfo  {journal} {Phys.
  Rev. X}\ }\textbf {\bibinfo {volume} {8}},\ \bibinfo {pages} {021052}
  (\bibinfo {year} {2018})}\BibitemShut {NoStop}%
\bibitem [{\citenamefont {Marshall}\ \emph {et~al.}(2003)\citenamefont
  {Marshall}, \citenamefont {Simon}, \citenamefont {Penrose},\ and\
  \citenamefont {Bouwmeester}}]{marshall_towards_2003}%
  \BibitemOpen
  \bibfield  {author} {\bibinfo {author} {\bibfnamefont {W.}~\bibnamefont
  {Marshall}}, \bibinfo {author} {\bibfnamefont {C.}~\bibnamefont {Simon}},
  \bibinfo {author} {\bibfnamefont {R.}~\bibnamefont {Penrose}}, \ and\
  \bibinfo {author} {\bibfnamefont {D.}~\bibnamefont {Bouwmeester}},\ }\href
  {\doibase 10.1103/PhysRevLett.91.130401} {\bibfield  {journal} {\bibinfo
  {journal} {Phys. Rev. Lett.}\ }\textbf {\bibinfo {volume} {91}},\ \bibinfo
  {pages} {130401} (\bibinfo {year} {2003})}\BibitemShut {NoStop}%
\bibitem [{\citenamefont {Pikovski}\ \emph {et~al.}(2012)\citenamefont
  {Pikovski}, \citenamefont {Vanner}, \citenamefont {Aspelmeyer}, \citenamefont
  {Kim},\ and\ \citenamefont {Brukner}}]{pikovski_probing_2012}%
  \BibitemOpen
  \bibfield  {author} {\bibinfo {author} {\bibfnamefont {I.}~\bibnamefont
  {Pikovski}}, \bibinfo {author} {\bibfnamefont {M.~R.}\ \bibnamefont
  {Vanner}}, \bibinfo {author} {\bibfnamefont {M.}~\bibnamefont {Aspelmeyer}},
  \bibinfo {author} {\bibfnamefont {M.~S.}\ \bibnamefont {Kim}}, \ and\
  \bibinfo {author} {\bibfnamefont {C.}~\bibnamefont {Brukner}},\ }\href
  {\doibase 10.1038/nphys2262} {\bibfield  {journal} {\bibinfo  {journal} {Nat
  Phys}\ }\textbf {\bibinfo {volume} {8}},\ \bibinfo {pages} {393} (\bibinfo
  {year} {2012})}\BibitemShut {NoStop}%
\bibitem [{\citenamefont {Li}\ \emph {et~al.}(2017)\citenamefont {Li},
  \citenamefont {Steane}, \citenamefont {Bedingham},\ and\ \citenamefont
  {Briggs}}]{li_detecting_2017}%
  \BibitemOpen
  \bibfield  {author} {\bibinfo {author} {\bibfnamefont {Y.}~\bibnamefont
  {Li}}, \bibinfo {author} {\bibfnamefont {A.~M.}\ \bibnamefont {Steane}},
  \bibinfo {author} {\bibfnamefont {D.}~\bibnamefont {Bedingham}}, \ and\
  \bibinfo {author} {\bibfnamefont {G.~A.~D.}\ \bibnamefont {Briggs}},\ }\href
  {\doibase 10.1103/PhysRevA.95.032112} {\bibfield  {journal} {\bibinfo
  {journal} {Physical Review A}\ }\textbf {\bibinfo {volume} {95}} (\bibinfo
  {year} {2017}),\ 10.1103/PhysRevA.95.032112},\ \bibinfo {note} {arXiv:
  1605.01881}\BibitemShut {NoStop}%
\bibitem [{\citenamefont {Nimmrichter}\ \emph {et~al.}(2014)\citenamefont
  {Nimmrichter}, \citenamefont {Hornberger},\ and\ \citenamefont
  {Hammerer}}]{nimmrichter_optomechanical_2014}%
  \BibitemOpen
  \bibfield  {author} {\bibinfo {author} {\bibfnamefont {S.}~\bibnamefont
  {Nimmrichter}}, \bibinfo {author} {\bibfnamefont {K.}~\bibnamefont
  {Hornberger}}, \ and\ \bibinfo {author} {\bibfnamefont {K.}~\bibnamefont
  {Hammerer}},\ }\href {\doibase 10.1103/PhysRevLett.113.020405} {\bibfield
  {journal} {\bibinfo  {journal} {Physical Review Letters}\ }\textbf {\bibinfo
  {volume} {113}} (\bibinfo {year} {2014}),\ 10.1103/PhysRevLett.113.020405},\
  \bibinfo {note} {arXiv: 1405.2868}\BibitemShut {NoStop}%
\bibitem [{\citenamefont {Bassi}\ \emph {et~al.}(2017)\citenamefont {Bassi},
  \citenamefont {Grossardt},\ and\ \citenamefont
  {Ulbricht}}]{bassi_gravitational_2017}%
  \BibitemOpen
  \bibfield  {author} {\bibinfo {author} {\bibfnamefont {A.}~\bibnamefont
  {Bassi}}, \bibinfo {author} {\bibfnamefont {A.}~\bibnamefont {Grossardt}}, \
  and\ \bibinfo {author} {\bibfnamefont {H.}~\bibnamefont {Ulbricht}},\ }\href
  {\doibase 10.1088/1361-6382/aa864f} {\bibfield  {journal} {\bibinfo
  {journal} {Class. Quantum Grav.}\ }\textbf {\bibinfo {volume} {34}},\
  \bibinfo {pages} {193002} (\bibinfo {year} {2017})}\BibitemShut {NoStop}%
\bibitem [{\citenamefont {Wallquist}\ \emph {et~al.}(2009)\citenamefont
  {Wallquist}, \citenamefont {Hammerer}, \citenamefont {Rabl}, \citenamefont
  {Lukin},\ and\ \citenamefont {Zoller}}]{wallquist_hybrid_2009}%
  \BibitemOpen
  \bibfield  {author} {\bibinfo {author} {\bibfnamefont {M.}~\bibnamefont
  {Wallquist}}, \bibinfo {author} {\bibfnamefont {K.}~\bibnamefont {Hammerer}},
  \bibinfo {author} {\bibfnamefont {P.}~\bibnamefont {Rabl}}, \bibinfo {author}
  {\bibfnamefont {M.}~\bibnamefont {Lukin}}, \ and\ \bibinfo {author}
  {\bibfnamefont {P.}~\bibnamefont {Zoller}},\ }\href {\doibase
  10.1088/0031-8949/2009/T137/014001} {\bibfield  {journal} {\bibinfo
  {journal} {Phys. Scr.}\ }\textbf {\bibinfo {volume} {2009}},\ \bibinfo
  {pages} {014001} (\bibinfo {year} {2009})}\BibitemShut {NoStop}%
\bibitem [{\citenamefont {Didier}\ \emph {et~al.}(2014)\citenamefont {Didier},
  \citenamefont {Pugnetti}, \citenamefont {Blanter},\ and\ \citenamefont
  {Fazio}}]{didier_quantum_2014}%
  \BibitemOpen
  \bibfield  {author} {\bibinfo {author} {\bibfnamefont {N.}~\bibnamefont
  {Didier}}, \bibinfo {author} {\bibfnamefont {S.}~\bibnamefont {Pugnetti}},
  \bibinfo {author} {\bibfnamefont {Y.~M.}\ \bibnamefont {Blanter}}, \ and\
  \bibinfo {author} {\bibfnamefont {R.}~\bibnamefont {Fazio}},\ }\href
  {\doibase 10.1016/j.ssc.2014.02.029} {\bibfield  {journal} {\bibinfo
  {journal} {Solid State Communications}\ }\bibinfo {series} {{SI}: {Spin}
  {Mechanics}},\ \textbf {\bibinfo {volume} {198}},\ \bibinfo {pages} {61}
  (\bibinfo {year} {2014})}\BibitemShut {NoStop}%
\bibitem [{\citenamefont {Gilchrist}\ \emph {et~al.}(2004)\citenamefont
  {Gilchrist}, \citenamefont {Nemoto}, \citenamefont {Munro}, \citenamefont
  {Ralph}, \citenamefont {Glancy}, \citenamefont {Braunstein},\ and\
  \citenamefont {Milburn}}]{gilchrist_schrodinger_2004}%
  \BibitemOpen
  \bibfield  {author} {\bibinfo {author} {\bibfnamefont {A.}~\bibnamefont
  {Gilchrist}}, \bibinfo {author} {\bibfnamefont {K.}~\bibnamefont {Nemoto}},
  \bibinfo {author} {\bibfnamefont {W.~J.}\ \bibnamefont {Munro}}, \bibinfo
  {author} {\bibfnamefont {T.~C.}\ \bibnamefont {Ralph}}, \bibinfo {author}
  {\bibfnamefont {S.}~\bibnamefont {Glancy}}, \bibinfo {author} {\bibfnamefont
  {S.~L.}\ \bibnamefont {Braunstein}}, \ and\ \bibinfo {author} {\bibfnamefont
  {G.~J.}\ \bibnamefont {Milburn}},\ }\href {\doibase
  10.1088/1464-4266/6/8/032} {\bibfield  {journal} {\bibinfo  {journal} {J.
  Opt. B: Quantum Semiclass. Opt.}\ }\textbf {\bibinfo {volume} {6}},\ \bibinfo
  {pages} {S828} (\bibinfo {year} {2004})}\BibitemShut {NoStop}%
\bibitem [{\citenamefont {Leghtas}\ \emph {et~al.}(2013)\citenamefont
  {Leghtas}, \citenamefont {Kirchmair}, \citenamefont {Vlastakis},
  \citenamefont {Schoelkopf}, \citenamefont {Devoret},\ and\ \citenamefont
  {Mirrahimi}}]{leghtas_hardware-efficient_2013}%
  \BibitemOpen
  \bibfield  {author} {\bibinfo {author} {\bibfnamefont {Z.}~\bibnamefont
  {Leghtas}}, \bibinfo {author} {\bibfnamefont {G.}~\bibnamefont {Kirchmair}},
  \bibinfo {author} {\bibfnamefont {B.}~\bibnamefont {Vlastakis}}, \bibinfo
  {author} {\bibfnamefont {R.~J.}\ \bibnamefont {Schoelkopf}}, \bibinfo
  {author} {\bibfnamefont {M.~H.}\ \bibnamefont {Devoret}}, \ and\ \bibinfo
  {author} {\bibfnamefont {M.}~\bibnamefont {Mirrahimi}},\ }\href {\doibase
  10.1103/PhysRevLett.111.120501} {\bibfield  {journal} {\bibinfo  {journal}
  {Phys. Rev. Lett.}\ }\textbf {\bibinfo {volume} {111}},\ \bibinfo {pages}
  {120501} (\bibinfo {year} {2013})}\BibitemShut {NoStop}%
\bibitem [{\citenamefont {Regal}\ and\ \citenamefont
  {Lehnert}(2011)}]{regal_cavity_2011}%
  \BibitemOpen
  \bibfield  {author} {\bibinfo {author} {\bibfnamefont {C.~A.}\ \bibnamefont
  {Regal}}\ and\ \bibinfo {author} {\bibfnamefont {K.~W.}\ \bibnamefont
  {Lehnert}},\ }\href {\doibase 10.1088/1742-6596/264/1/012025} {\bibfield
  {journal} {\bibinfo  {journal} {J. Phys.: Conf. Ser.}\ }\textbf {\bibinfo
  {volume} {264}},\ \bibinfo {pages} {012025} (\bibinfo {year}
  {2011})}\BibitemShut {NoStop}%
\bibitem [{\citenamefont {Harris}\ \emph {et~al.}(2012)\citenamefont {Harris},
  \citenamefont {Andersen}, \citenamefont {Knittel},\ and\ \citenamefont
  {Bowen}}]{harris_feedback-enhanced_2012}%
  \BibitemOpen
  \bibfield  {author} {\bibinfo {author} {\bibfnamefont {G.}~\bibnamefont
  {Harris}}, \bibinfo {author} {\bibfnamefont {U.}~\bibnamefont {Andersen}},
  \bibinfo {author} {\bibfnamefont {J.}~\bibnamefont {Knittel}}, \ and\
  \bibinfo {author} {\bibfnamefont {W.}~\bibnamefont {Bowen}},\ }\href
  {\doibase 10.1103/PhysRevA.85.061802} {\bibfield  {journal} {\bibinfo
  {journal} {Phys. Rev. A}\ }\textbf {\bibinfo {volume} {85}},\ \bibinfo
  {pages} {061802} (\bibinfo {year} {2012})}\BibitemShut {NoStop}%
\bibitem [{\citenamefont {Khosla}\ \emph {et~al.}(2017)\citenamefont {Khosla},
  \citenamefont {Brawley}, \citenamefont {Vanner},\ and\ \citenamefont
  {Bowen}}]{khosla_quantum_2017}%
  \BibitemOpen
  \bibfield  {author} {\bibinfo {author} {\bibfnamefont {K.~E.}\ \bibnamefont
  {Khosla}}, \bibinfo {author} {\bibfnamefont {G.~A.}\ \bibnamefont {Brawley}},
  \bibinfo {author} {\bibfnamefont {M.~R.}\ \bibnamefont {Vanner}}, \ and\
  \bibinfo {author} {\bibfnamefont {W.~P.}\ \bibnamefont {Bowen}},\ }\href
  {\doibase 10.1364/OPTICA.4.001382} {\bibfield  {journal} {\bibinfo  {journal}
  {Optica, OPTICA}\ }\textbf {\bibinfo {volume} {4}},\ \bibinfo {pages} {1382}
  (\bibinfo {year} {2017})}\BibitemShut {NoStop}%
\bibitem [{\citenamefont {Forstner}\ \emph {et~al.}(2014)\citenamefont
  {Forstner}, \citenamefont {Sheridan}, \citenamefont {Knittel}, \citenamefont
  {Humphreys}, \citenamefont {Brawley}, \citenamefont {Rubinsztein-Dunlop},\
  and\ \citenamefont {Bowen}}]{forstner_ultrasensitive_2014}%
  \BibitemOpen
  \bibfield  {author} {\bibinfo {author} {\bibfnamefont {S.}~\bibnamefont
  {Forstner}}, \bibinfo {author} {\bibfnamefont {E.}~\bibnamefont {Sheridan}},
  \bibinfo {author} {\bibfnamefont {J.}~\bibnamefont {Knittel}}, \bibinfo
  {author} {\bibfnamefont {C.~L.}\ \bibnamefont {Humphreys}}, \bibinfo {author}
  {\bibfnamefont {G.~A.}\ \bibnamefont {Brawley}}, \bibinfo {author}
  {\bibfnamefont {H.}~\bibnamefont {Rubinsztein-Dunlop}}, \ and\ \bibinfo
  {author} {\bibfnamefont {W.~P.}\ \bibnamefont {Bowen}},\ }\href {\doibase
  10.1002/adma.201401144} {\bibfield  {journal} {\bibinfo  {journal} {Adv.
  Mater.}\ }\textbf {\bibinfo {volume} {26}},\ \bibinfo {pages} {6348}
  (\bibinfo {year} {2014})}\BibitemShut {NoStop}%
\bibitem [{\citenamefont {Toscano}\ \emph {et~al.}(2006)\citenamefont
  {Toscano}, \citenamefont {Dalvit}, \citenamefont {Davidovich},\ and\
  \citenamefont {Zurek}}]{toscano_sub-planck_2006}%
  \BibitemOpen
  \bibfield  {author} {\bibinfo {author} {\bibfnamefont {F.}~\bibnamefont
  {Toscano}}, \bibinfo {author} {\bibfnamefont {D.~A.~R.}\ \bibnamefont
  {Dalvit}}, \bibinfo {author} {\bibfnamefont {L.}~\bibnamefont {Davidovich}},
  \ and\ \bibinfo {author} {\bibfnamefont {W.~H.}\ \bibnamefont {Zurek}},\
  }\href {\doibase 10.1103/PhysRevA.73.023803} {\bibfield  {journal} {\bibinfo
  {journal} {Phys. Rev. A}\ }\textbf {\bibinfo {volume} {73}},\ \bibinfo
  {pages} {023803} (\bibinfo {year} {2006})}\BibitemShut {NoStop}%
\bibitem [{\citenamefont {Palomaki}\ \emph
  {et~al.}(2013{\natexlab{b}})\citenamefont {Palomaki}, \citenamefont {Harlow},
  \citenamefont {Teufel}, \citenamefont {Simmonds},\ and\ \citenamefont
  {Lehnert}}]{palomaki_coherent_2013}%
  \BibitemOpen
  \bibfield  {author} {\bibinfo {author} {\bibfnamefont {T.~A.}\ \bibnamefont
  {Palomaki}}, \bibinfo {author} {\bibfnamefont {J.~W.}\ \bibnamefont
  {Harlow}}, \bibinfo {author} {\bibfnamefont {J.~D.}\ \bibnamefont {Teufel}},
  \bibinfo {author} {\bibfnamefont {R.~W.}\ \bibnamefont {Simmonds}}, \ and\
  \bibinfo {author} {\bibfnamefont {K.~W.}\ \bibnamefont {Lehnert}},\ }\href
  {\doibase 10.1038/nature11915} {\bibfield  {journal} {\bibinfo  {journal}
  {Nature}\ }\textbf {\bibinfo {volume} {495}},\ \bibinfo {pages} {210}
  (\bibinfo {year} {2013}{\natexlab{b}})}\BibitemShut {NoStop}%
\bibitem [{\citenamefont {Bennett}\ \emph {et~al.}(2016)\citenamefont
  {Bennett}, \citenamefont {Khosla}, \citenamefont {Madsen}, \citenamefont
  {Vanner}, \citenamefont {Rubinsztein-Dunlop},\ and\ \citenamefont
  {Bowen}}]{bennett_quantum_2016}%
  \BibitemOpen
  \bibfield  {author} {\bibinfo {author} {\bibfnamefont {J.~S.}\ \bibnamefont
  {Bennett}}, \bibinfo {author} {\bibfnamefont {K.}~\bibnamefont {Khosla}},
  \bibinfo {author} {\bibfnamefont {L.~S.}\ \bibnamefont {Madsen}}, \bibinfo
  {author} {\bibfnamefont {M.~R.}\ \bibnamefont {Vanner}}, \bibinfo {author}
  {\bibfnamefont {H.}~\bibnamefont {Rubinsztein-Dunlop}}, \ and\ \bibinfo
  {author} {\bibfnamefont {W.~P.}\ \bibnamefont {Bowen}},\ }\href {\doibase
  10.1088/1367-2630/18/5/053030} {\bibfield  {journal} {\bibinfo  {journal}
  {New J. Phys.}\ }\textbf {\bibinfo {volume} {18}},\ \bibinfo {pages} {053030}
  (\bibinfo {year} {2016})}\BibitemShut {NoStop}%
\bibitem [{\citenamefont {Caves}\ and\ \citenamefont
  {Milburn}(1987)}]{caves_quantum-mechanical_1987}%
  \BibitemOpen
  \bibfield  {author} {\bibinfo {author} {\bibfnamefont {C.~M.}\ \bibnamefont
  {Caves}}\ and\ \bibinfo {author} {\bibfnamefont {G.~J.}\ \bibnamefont
  {Milburn}},\ }\href {\doibase 10.1103/PhysRevA.36.5543} {\bibfield  {journal}
  {\bibinfo  {journal} {Phys. Rev. A}\ }\textbf {\bibinfo {volume} {36}},\
  \bibinfo {pages} {5543} (\bibinfo {year} {1987})}\BibitemShut {NoStop}%
\bibitem [{\citenamefont {Schneider}\ \emph {et~al.}(2012)\citenamefont
  {Schneider}, \citenamefont {Etaki}, \citenamefont {van~der Zant},\ and\
  \citenamefont {Steele}}]{schneider_coupling_2012}%
  \BibitemOpen
  \bibfield  {author} {\bibinfo {author} {\bibfnamefont {B.~H.}\ \bibnamefont
  {Schneider}}, \bibinfo {author} {\bibfnamefont {S.}~\bibnamefont {Etaki}},
  \bibinfo {author} {\bibfnamefont {H.~S.~J.}\ \bibnamefont {van~der Zant}}, \
  and\ \bibinfo {author} {\bibfnamefont {G.~A.}\ \bibnamefont {Steele}},\
  }\href {\doibase 10.1038/srep00599} {\bibfield  {journal} {\bibinfo
  {journal} {Sci Rep}\ }\textbf {\bibinfo {volume} {2}} (\bibinfo {year}
  {2012}),\ 10.1038/srep00599}\BibitemShut {NoStop}%
\bibitem [{\citenamefont {Cleuziou}\ \emph {et~al.}(2006)\citenamefont
  {Cleuziou}, \citenamefont {Wernsdorfer}, \citenamefont {Bouchiat},
  \citenamefont {Ondarcuhu},\ and\ \citenamefont
  {Monthioux}}]{cleuziou_carbon_2006}%
  \BibitemOpen
  \bibfield  {author} {\bibinfo {author} {\bibfnamefont {J.-P.}\ \bibnamefont
  {Cleuziou}}, \bibinfo {author} {\bibfnamefont {W.}~\bibnamefont
  {Wernsdorfer}}, \bibinfo {author} {\bibfnamefont {V.}~\bibnamefont
  {Bouchiat}}, \bibinfo {author} {\bibfnamefont {T.}~\bibnamefont {Ondarcuhu}},
  \ and\ \bibinfo {author} {\bibfnamefont {M.}~\bibnamefont {Monthioux}},\
  }\href {\doibase 10.1038/nnano.2006.54} {\bibfield  {journal} {\bibinfo
  {journal} {Nat Nano}\ }\textbf {\bibinfo {volume} {1}},\ \bibinfo {pages}
  {53} (\bibinfo {year} {2006})}\BibitemShut {NoStop}%
\bibitem [{\citenamefont {Arcizet}\ \emph {et~al.}(2011)\citenamefont
  {Arcizet}, \citenamefont {Jacques}, \citenamefont {Siria}, \citenamefont
  {Poncharal}, \citenamefont {Vincent},\ and\ \citenamefont
  {Seidelin}}]{arcizet_single_2011}%
  \BibitemOpen
  \bibfield  {author} {\bibinfo {author} {\bibfnamefont {O.}~\bibnamefont
  {Arcizet}}, \bibinfo {author} {\bibfnamefont {V.}~\bibnamefont {Jacques}},
  \bibinfo {author} {\bibfnamefont {A.}~\bibnamefont {Siria}}, \bibinfo
  {author} {\bibfnamefont {P.}~\bibnamefont {Poncharal}}, \bibinfo {author}
  {\bibfnamefont {P.}~\bibnamefont {Vincent}}, \ and\ \bibinfo {author}
  {\bibfnamefont {S.}~\bibnamefont {Seidelin}},\ }\href {\doibase
  10.1038/nphys2070} {\bibfield  {journal} {\bibinfo  {journal} {Nature
  Physics}\ }\textbf {\bibinfo {volume} {7}},\ \bibinfo {pages} {879} (\bibinfo
  {year} {2011})}\BibitemShut {NoStop}%
\bibitem [{\citenamefont {Brawley}\ \emph {et~al.}(2016)\citenamefont
  {Brawley}, \citenamefont {Vanner}, \citenamefont {Larsen}, \citenamefont
  {Schmid}, \citenamefont {Boisen},\ and\ \citenamefont
  {Bowen}}]{brawley_nonlinear_2016}%
  \BibitemOpen
  \bibfield  {author} {\bibinfo {author} {\bibfnamefont {G.~A.}\ \bibnamefont
  {Brawley}}, \bibinfo {author} {\bibfnamefont {M.~R.}\ \bibnamefont {Vanner}},
  \bibinfo {author} {\bibfnamefont {P.~E.}\ \bibnamefont {Larsen}}, \bibinfo
  {author} {\bibfnamefont {S.}~\bibnamefont {Schmid}}, \bibinfo {author}
  {\bibfnamefont {A.}~\bibnamefont {Boisen}}, \ and\ \bibinfo {author}
  {\bibfnamefont {W.~P.}\ \bibnamefont {Bowen}},\ }\href {\doibase
  10.1038/ncomms10988} {\bibfield  {journal} {\bibinfo  {journal} {Nature
  Communications}\ }\textbf {\bibinfo {volume} {7}},\ \bibinfo {pages} {10988}
  (\bibinfo {year} {2016})}\BibitemShut {NoStop}%
\bibitem [{\citenamefont {Brune}\ \emph {et~al.}(1992)\citenamefont {Brune},
  \citenamefont {Haroche}, \citenamefont {Raimond}, \citenamefont
  {Davidovich},\ and\ \citenamefont {Zagury}}]{brune_manipulation_1992}%
  \BibitemOpen
  \bibfield  {author} {\bibinfo {author} {\bibfnamefont {M.}~\bibnamefont
  {Brune}}, \bibinfo {author} {\bibfnamefont {S.}~\bibnamefont {Haroche}},
  \bibinfo {author} {\bibfnamefont {J.~M.}\ \bibnamefont {Raimond}}, \bibinfo
  {author} {\bibfnamefont {L.}~\bibnamefont {Davidovich}}, \ and\ \bibinfo
  {author} {\bibfnamefont {N.}~\bibnamefont {Zagury}},\ }\href {\doibase
  10.1103/PhysRevA.45.5193} {\bibfield  {journal} {\bibinfo  {journal} {Phys.
  Rev. A}\ }\textbf {\bibinfo {volume} {45}},\ \bibinfo {pages} {5193}
  (\bibinfo {year} {1992})}\BibitemShut {NoStop}%
\bibitem [{\citenamefont {Opatrny}\ \emph {et~al.}(2000)\citenamefont
  {Opatrny}, \citenamefont {Kurizki},\ and\ \citenamefont
  {Welsch}}]{opatrny_improvement_2000}%
  \BibitemOpen
  \bibfield  {author} {\bibinfo {author} {\bibfnamefont {T.}~\bibnamefont
  {Opatrny}}, \bibinfo {author} {\bibfnamefont {G.}~\bibnamefont {Kurizki}}, \
  and\ \bibinfo {author} {\bibfnamefont {D.-G.}\ \bibnamefont {Welsch}},\
  }\href {\doibase 10.1103/PhysRevA.61.032302} {\bibfield  {journal} {\bibinfo
  {journal} {Phys. Rev. A}\ }\textbf {\bibinfo {volume} {61}},\ \bibinfo
  {pages} {032302} (\bibinfo {year} {2000})}\BibitemShut {NoStop}%
\bibitem [{\citenamefont {Reed}\ \emph {et~al.}(2017)\citenamefont {Reed},
  \citenamefont {Mayer}, \citenamefont {Teufel}, \citenamefont {Burkhart},
  \citenamefont {Pfaff}, \citenamefont {Reagor}, \citenamefont {Sletten},
  \citenamefont {Ma}, \citenamefont {Schoelkopf}, \citenamefont {Knill},\ and\
  \citenamefont {Lehnert}}]{reed_faithful_2017}%
  \BibitemOpen
  \bibfield  {author} {\bibinfo {author} {\bibfnamefont {A.~P.}\ \bibnamefont
  {Reed}}, \bibinfo {author} {\bibfnamefont {K.~H.}\ \bibnamefont {Mayer}},
  \bibinfo {author} {\bibfnamefont {J.~D.}\ \bibnamefont {Teufel}}, \bibinfo
  {author} {\bibfnamefont {L.~D.}\ \bibnamefont {Burkhart}}, \bibinfo {author}
  {\bibfnamefont {W.}~\bibnamefont {Pfaff}}, \bibinfo {author} {\bibfnamefont
  {M.}~\bibnamefont {Reagor}}, \bibinfo {author} {\bibfnamefont
  {L.}~\bibnamefont {Sletten}}, \bibinfo {author} {\bibfnamefont
  {X.}~\bibnamefont {Ma}}, \bibinfo {author} {\bibfnamefont {R.~J.}\
  \bibnamefont {Schoelkopf}}, \bibinfo {author} {\bibfnamefont
  {E.}~\bibnamefont {Knill}}, \ and\ \bibinfo {author} {\bibfnamefont {K.~W.}\
  \bibnamefont {Lehnert}},\ }\href {\doibase 10.1038/nphys4251} {\bibfield
  {journal} {\bibinfo  {journal} {Nature Physics}\ }\textbf {\bibinfo {volume}
  {13}},\ \bibinfo {pages} {1163} (\bibinfo {year} {2017})}\BibitemShut
  {NoStop}%
\bibitem [{\citenamefont {Khosla}\ \emph {et~al.}(2013)\citenamefont {Khosla},
  \citenamefont {Vanner}, \citenamefont {Bowen},\ and\ \citenamefont
  {Milburn}}]{khosla_quantum_2013}%
  \BibitemOpen
  \bibfield  {author} {\bibinfo {author} {\bibfnamefont {K.~E.}\ \bibnamefont
  {Khosla}}, \bibinfo {author} {\bibfnamefont {M.~R.}\ \bibnamefont {Vanner}},
  \bibinfo {author} {\bibfnamefont {W.~P.}\ \bibnamefont {Bowen}}, \ and\
  \bibinfo {author} {\bibfnamefont {G.~J.}\ \bibnamefont {Milburn}},\ }\href
  {\doibase 10.1088/1367-2630/15/4/043025} {\bibfield  {journal} {\bibinfo
  {journal} {New J. Phys.}\ }\textbf {\bibinfo {volume} {15}},\ \bibinfo
  {pages} {043025} (\bibinfo {year} {2013})}\BibitemShut {NoStop}%
\bibitem [{\citenamefont {Hofheinz}\ \emph {et~al.}(2009)\citenamefont
  {Hofheinz}, \citenamefont {Wang}, \citenamefont {Ansmann}, \citenamefont
  {Bialczak}, \citenamefont {Lucero}, \citenamefont {Neeley}, \citenamefont
  {O'Connell}, \citenamefont {Sank}, \citenamefont {Wenner}, \citenamefont
  {Martinis},\ and\ \citenamefont {Cleland}}]{hofheinz_synthesizing_2009}%
  \BibitemOpen
  \bibfield  {author} {\bibinfo {author} {\bibfnamefont {M.}~\bibnamefont
  {Hofheinz}}, \bibinfo {author} {\bibfnamefont {H.}~\bibnamefont {Wang}},
  \bibinfo {author} {\bibfnamefont {M.}~\bibnamefont {Ansmann}}, \bibinfo
  {author} {\bibfnamefont {R.~C.}\ \bibnamefont {Bialczak}}, \bibinfo {author}
  {\bibfnamefont {E.}~\bibnamefont {Lucero}}, \bibinfo {author} {\bibfnamefont
  {M.}~\bibnamefont {Neeley}}, \bibinfo {author} {\bibfnamefont {A.~D.}\
  \bibnamefont {O'Connell}}, \bibinfo {author} {\bibfnamefont {D.}~\bibnamefont
  {Sank}}, \bibinfo {author} {\bibfnamefont {J.}~\bibnamefont {Wenner}},
  \bibinfo {author} {\bibfnamefont {J.~M.}\ \bibnamefont {Martinis}}, \ and\
  \bibinfo {author} {\bibfnamefont {A.~N.}\ \bibnamefont {Cleland}},\ }\href
  {\doibase 10.1038/nature08005} {\bibfield  {journal} {\bibinfo  {journal}
  {Nature}\ }\textbf {\bibinfo {volume} {459}},\ \bibinfo {pages} {546}
  (\bibinfo {year} {2009})}\BibitemShut {NoStop}%
\bibitem [{\citenamefont {Berdova}\ \emph {et~al.}(2013)\citenamefont
  {Berdova}, \citenamefont {Cho}, \citenamefont {Pirkkalainen}, \citenamefont
  {Sulkko}, \citenamefont {Song}, \citenamefont {Hakonen},\ and\ \citenamefont
  {Sillanpaa}}]{berdova_micromanipulation_2013}%
  \BibitemOpen
  \bibfield  {author} {\bibinfo {author} {\bibfnamefont {M.}~\bibnamefont
  {Berdova}}, \bibinfo {author} {\bibfnamefont {S.~U.}\ \bibnamefont {Cho}},
  \bibinfo {author} {\bibfnamefont {J.-M.}\ \bibnamefont {Pirkkalainen}},
  \bibinfo {author} {\bibfnamefont {J.}~\bibnamefont {Sulkko}}, \bibinfo
  {author} {\bibfnamefont {X.}~\bibnamefont {Song}}, \bibinfo {author}
  {\bibfnamefont {P.~J.}\ \bibnamefont {Hakonen}}, \ and\ \bibinfo {author}
  {\bibfnamefont {M.~A.}\ \bibnamefont {Sillanpaa}},\ }\href {\doibase
  10.1088/0960-1317/23/12/125024} {\bibfield  {journal} {\bibinfo  {journal}
  {Journal of Micromechanics and Microengineering}\ }\textbf {\bibinfo {volume}
  {23}},\ \bibinfo {pages} {125024} (\bibinfo {year} {2013})},\ \bibinfo {note}
  {arXiv: 1307.1619}\BibitemShut {NoStop}%
\bibitem [{\citenamefont {Vogel}\ \emph {et~al.}(1993)\citenamefont {Vogel},
  \citenamefont {Akulin},\ and\ \citenamefont {Schleich}}]{vogel_quantum_1993}%
  \BibitemOpen
  \bibfield  {author} {\bibinfo {author} {\bibfnamefont {K.}~\bibnamefont
  {Vogel}}, \bibinfo {author} {\bibfnamefont {V.~M.}\ \bibnamefont {Akulin}}, \
  and\ \bibinfo {author} {\bibfnamefont {W.~P.}\ \bibnamefont {Schleich}},\
  }\href {\doibase 10.1103/PhysRevLett.71.1816} {\bibfield  {journal} {\bibinfo
   {journal} {Phys. Rev. Lett.}\ }\textbf {\bibinfo {volume} {71}},\ \bibinfo
  {pages} {1816} (\bibinfo {year} {1993})}\BibitemShut {NoStop}%
\bibitem [{\citenamefont {Makhlin}\ \emph {et~al.}(2001)\citenamefont
  {Makhlin}, \citenamefont {Schon},\ and\ \citenamefont
  {Shnirman}}]{makhlin_quantum-state_2001}%
  \BibitemOpen
  \bibfield  {author} {\bibinfo {author} {\bibfnamefont {Y.}~\bibnamefont
  {Makhlin}}, \bibinfo {author} {\bibfnamefont {G.}~\bibnamefont {Schon}}, \
  and\ \bibinfo {author} {\bibfnamefont {A.}~\bibnamefont {Shnirman}},\ }\href
  {\doibase 10.1103/RevModPhys.73.357} {\bibfield  {journal} {\bibinfo
  {journal} {Rev. Mod. Phys.}\ }\textbf {\bibinfo {volume} {73}},\ \bibinfo
  {pages} {357} (\bibinfo {year} {2001})}\BibitemShut {NoStop}%
\bibitem [{\citenamefont {Wiseman}\ and\ \citenamefont
  {Milburn}(2009)}]{wiseman_quantum_2009}%
  \BibitemOpen
  \bibfield  {author} {\bibinfo {author} {\bibfnamefont {H.~M.}\ \bibnamefont
  {Wiseman}}\ and\ \bibinfo {author} {\bibfnamefont {G.~J.}\ \bibnamefont
  {Milburn}},\ }\href@noop {} {\emph {\bibinfo {title} {Quantum {Measurement}
  and {Control}}}}\ (\bibinfo  {publisher} {Cambridge University Press},\
  \bibinfo {address} {Cambridge},\ \bibinfo {year} {2009})\BibitemShut
  {NoStop}%
\bibitem [{\citenamefont {Vanner}\ \emph {et~al.}(2011)\citenamefont {Vanner},
  \citenamefont {Pikovski}, \citenamefont {Cole}, \citenamefont {Kim},
  \citenamefont {Brukner}, \citenamefont {Hammerer}, \citenamefont {Milburn},\
  and\ \citenamefont {Aspelmeyer}}]{vanner_pulsed_2011}%
  \BibitemOpen
  \bibfield  {author} {\bibinfo {author} {\bibfnamefont {M.~R.}\ \bibnamefont
  {Vanner}}, \bibinfo {author} {\bibfnamefont {I.}~\bibnamefont {Pikovski}},
  \bibinfo {author} {\bibfnamefont {G.~D.}\ \bibnamefont {Cole}}, \bibinfo
  {author} {\bibfnamefont {M.~S.}\ \bibnamefont {Kim}}, \bibinfo {author}
  {\bibfnamefont {C.}~\bibnamefont {Brukner}}, \bibinfo {author} {\bibfnamefont
  {K.}~\bibnamefont {Hammerer}}, \bibinfo {author} {\bibfnamefont {G.~J.}\
  \bibnamefont {Milburn}}, \ and\ \bibinfo {author} {\bibfnamefont
  {M.}~\bibnamefont {Aspelmeyer}},\ }\href {\doibase 10.1073/pnas.1105098108}
  {\bibfield  {journal} {\bibinfo  {journal} {PNAS}\ }\textbf {\bibinfo
  {volume} {108}},\ \bibinfo {pages} {16182} (\bibinfo {year}
  {2011})}\BibitemShut {NoStop}%
\bibitem [{\citenamefont {Wiseman}(1996)}]{wiseman_quantum_1996}%
  \BibitemOpen
  \bibfield  {author} {\bibinfo {author} {\bibfnamefont {H.~M.}\ \bibnamefont
  {Wiseman}},\ }\href {\doibase 10.1088/1355-5111/8/1/015} {\bibfield
  {journal} {\bibinfo  {journal} {Quantum Semiclass. Opt.}\ }\textbf {\bibinfo
  {volume} {8}},\ \bibinfo {pages} {205} (\bibinfo {year} {1996})}\BibitemShut
  {NoStop}%
\bibitem [{\citenamefont {Vanner}\ \emph {et~al.}(2013)\citenamefont {Vanner},
  \citenamefont {Hofer}, \citenamefont {Cole},\ and\ \citenamefont
  {Aspelmeyer}}]{vanner_cooling-by-measurement_2013}%
  \BibitemOpen
  \bibfield  {author} {\bibinfo {author} {\bibfnamefont {M.~R.}\ \bibnamefont
  {Vanner}}, \bibinfo {author} {\bibfnamefont {J.}~\bibnamefont {Hofer}},
  \bibinfo {author} {\bibfnamefont {G.~D.}\ \bibnamefont {Cole}}, \ and\
  \bibinfo {author} {\bibfnamefont {M.}~\bibnamefont {Aspelmeyer}},\ }\href
  {\doibase 10.1038/ncomms3295} {\bibfield  {journal} {\bibinfo  {journal} {Nat
  Commun}\ }\textbf {\bibinfo {volume} {4}},\ \bibinfo {pages} {2295} (\bibinfo
  {year} {2013})}\BibitemShut {NoStop}%
\bibitem [{\citenamefont {Walls}\ and\ \citenamefont
  {Milburn}(2008)}]{walls_quantum_2008}%
  \BibitemOpen
  \bibfield  {author} {\bibinfo {author} {\bibfnamefont {D.~F.}\ \bibnamefont
  {Walls}}\ and\ \bibinfo {author} {\bibfnamefont {G.~J.}\ \bibnamefont
  {Milburn}},\ }\href {//www.springer.com/gb/book/9783540285731} {\emph
  {\bibinfo {title} {Quantum {Optics}}}},\ \bibinfo {edition} {2nd}\ ed.\
  (\bibinfo  {publisher} {Springer-Verlag},\ \bibinfo {address} {Berlin
  Heidelberg},\ \bibinfo {year} {2008})\BibitemShut {NoStop}%
\bibitem [{\citenamefont {Vanner}(2011)}]{vanner_selective_2011}%
  \BibitemOpen
  \bibfield  {author} {\bibinfo {author} {\bibfnamefont {M.~R.}\ \bibnamefont
  {Vanner}},\ }\href {\doibase 10.1103/PhysRevX.1.021011} {\bibfield  {journal}
  {\bibinfo  {journal} {Phys. Rev. X}\ }\textbf {\bibinfo {volume} {1}},\
  \bibinfo {pages} {021011} (\bibinfo {year} {2011})}\BibitemShut {NoStop}%
\bibitem [{\citenamefont {Clarke}\ and\ \citenamefont
  {Vanner}(2018)}]{clarke_growing_2018}%
  \BibitemOpen
  \bibfield  {author} {\bibinfo {author} {\bibfnamefont {J.}~\bibnamefont
  {Clarke}}\ and\ \bibinfo {author} {\bibfnamefont {M.~R.}\ \bibnamefont
  {Vanner}},\ }\href {http://arxiv.org/abs/1805.09334} {\bibfield  {journal}
  {\bibinfo  {journal} {arXiv:1805.09334 [cond-mat, physics:physics,
  physics:quant-ph]}\ } (\bibinfo {year} {2018})},\ \bibinfo {note} {arXiv:
  1805.09334}\BibitemShut {NoStop}%
\bibitem [{\citenamefont {Moser}\ \emph {et~al.}(2014)\citenamefont {Moser},
  \citenamefont {Eichler}, \citenamefont {Güttinger}, \citenamefont {Dykman},\
  and\ \citenamefont {Bachtold}}]{moser_nanotube_2014}%
  \BibitemOpen
  \bibfield  {author} {\bibinfo {author} {\bibfnamefont {J.}~\bibnamefont
  {Moser}}, \bibinfo {author} {\bibfnamefont {A.}~\bibnamefont {Eichler}},
  \bibinfo {author} {\bibfnamefont {J.}~\bibnamefont {Güttinger}}, \bibinfo
  {author} {\bibfnamefont {M.~I.}\ \bibnamefont {Dykman}}, \ and\ \bibinfo
  {author} {\bibfnamefont {A.}~\bibnamefont {Bachtold}},\ }\href {\doibase
  10.1038/nnano.2014.234} {\bibfield  {journal} {\bibinfo  {journal} {Nat
  Nano}\ }\textbf {\bibinfo {volume} {9}},\ \bibinfo {pages} {1007} (\bibinfo
  {year} {2014})}\BibitemShut {NoStop}%
\bibitem [{\citenamefont {Vogel}\ and\ \citenamefont
  {Risken}(1989)}]{vogel_determination_1989}%
  \BibitemOpen
  \bibfield  {author} {\bibinfo {author} {\bibfnamefont {K.}~\bibnamefont
  {Vogel}}\ and\ \bibinfo {author} {\bibfnamefont {H.}~\bibnamefont {Risken}},\
  }\href {\doibase 10.1103/PhysRevA.40.2847} {\bibfield  {journal} {\bibinfo
  {journal} {Phys. Rev. A}\ }\textbf {\bibinfo {volume} {40}},\ \bibinfo
  {pages} {2847} (\bibinfo {year} {1989})}\BibitemShut {NoStop}%
\bibitem [{\citenamefont {Smithey}\ \emph {et~al.}(1993)\citenamefont
  {Smithey}, \citenamefont {Beck}, \citenamefont {Raymer},\ and\ \citenamefont
  {Faridani}}]{smithey_measurement_1993}%
  \BibitemOpen
  \bibfield  {author} {\bibinfo {author} {\bibfnamefont {D.~T.}\ \bibnamefont
  {Smithey}}, \bibinfo {author} {\bibfnamefont {M.}~\bibnamefont {Beck}},
  \bibinfo {author} {\bibfnamefont {M.~G.}\ \bibnamefont {Raymer}}, \ and\
  \bibinfo {author} {\bibfnamefont {A.}~\bibnamefont {Faridani}},\ }\href
  {\doibase 10.1103/PhysRevLett.70.1244} {\bibfield  {journal} {\bibinfo
  {journal} {Phys. Rev. Lett.}\ }\textbf {\bibinfo {volume} {70}},\ \bibinfo
  {pages} {1244} (\bibinfo {year} {1993})}\BibitemShut {NoStop}%
\bibitem [{\citenamefont {Lvovsky}\ and\ \citenamefont
  {Raymer}(2009)}]{lvovsky_continuous-variable_2009}%
  \BibitemOpen
  \bibfield  {author} {\bibinfo {author} {\bibfnamefont {A.~I.}\ \bibnamefont
  {Lvovsky}}\ and\ \bibinfo {author} {\bibfnamefont {M.~G.}\ \bibnamefont
  {Raymer}},\ }\href {\doibase 10.1103/RevModPhys.81.299} {\bibfield  {journal}
  {\bibinfo  {journal} {Rev. Mod. Phys.}\ }\textbf {\bibinfo {volume} {81}},\
  \bibinfo {pages} {299} (\bibinfo {year} {2009})}\BibitemShut {NoStop}%
\bibitem [{\citenamefont {Yadin}\ \emph {et~al.}(2018)\citenamefont {Yadin},
  \citenamefont {Narasimhachar}, \citenamefont {Binder}, \citenamefont
  {Thompson}, \citenamefont {Gu},\ and\ \citenamefont
  {Kim}}]{yadin_operational_2018}%
  \BibitemOpen
  \bibfield  {author} {\bibinfo {author} {\bibfnamefont {B.}~\bibnamefont
  {Yadin}}, \bibinfo {author} {\bibfnamefont {V.}~\bibnamefont
  {Narasimhachar}}, \bibinfo {author} {\bibfnamefont {F.~C.}\ \bibnamefont
  {Binder}}, \bibinfo {author} {\bibfnamefont {J.}~\bibnamefont {Thompson}},
  \bibinfo {author} {\bibfnamefont {M.}~\bibnamefont {Gu}}, \ and\ \bibinfo
  {author} {\bibfnamefont {M.~S.}\ \bibnamefont {Kim}},\ }\href
  {http://arxiv.org/abs/1804.10190} {\bibfield  {journal} {\bibinfo  {journal}
  {arXiv:1804.10190 [quant-ph]}\ } (\bibinfo {year} {2018})},\ \bibinfo {note}
  {arXiv: 1804.10190}\BibitemShut {NoStop}%
\end{thebibliography}%

\end{document}